\documentclass[graybox]{svmult}

\usepackage{mathptmx}       
\usepackage{helvet}         
\usepackage{courier}        
\usepackage{type1cm}        
\usepackage{makeidx}         
\usepackage{graphicx}        
\usepackage[bottom]{footmisc}

\usepackage[comma,authoryear]{natbib}

\usepackage{subfig}
\usepackage{tabularx}
\usepackage{multirow}
\usepackage{rotating}

\usepackage{amstext, amsmath, amssymb}

\usepackage{hyperxmp}
\usepackage{hyperref}
\hypersetup{colorlinks=true, urlcolor=blue, citecolor=cyan}

\hypersetup{
    bookmarks=true,         
    unicode=true,          
    pdftoolbar=true,        
    pdfmenubar=true,        
    pdffitwindow=false,     
    pdfstartview={FitH},    
    pdftitle={Optimal Design of the Shiryaev--Roberts Chart: Give Your Shiryaev--Roberts a Headstart},    
    pdfauthor={Aleksey S. Polunchenko},     
    pdfsubject={Optimal Design of the Shiryaev--Roberts Chart: Give Your Shiryaev--Roberts a Headstart},   
    pdfcreator={Aleksey S. Polunchenko},   
    pdfproducer={MikTeX}, 
    pdfkeywords={Average Run Length, Fast Initial Response feature, Generalized Shiryaev--Roberts chart, Quality control, Quickest change-point detection, Sequential analysis}, 
    pdfnewwindow=true      
}

\usepackage{doi}
\usepackage[T1]{fontenc}

\renewcommand{\Pr}{\mathbb{P}} 
\DeclareMathOperator{\EV}{\mathbb{E}} 

\DeclareMathOperator{\ADD}{ADD}
\DeclareMathOperator{\ARL}{ARL}
\DeclareMathOperator{\SADD}{SADD}
\DeclareMathOperator{\STADD}{STADD}
\DeclareMathOperator{\RIADD}{RIADD}
\DeclareMathOperator{\IADD}{IADD}

\DeclareMathOperator*{\argmin}{arg\,min}

\newcommand{\T}{T}

\begin{document}

\title*{Optimal Design of the Shiryaev--Roberts Chart:\texorpdfstring{\\}{}
Give Your Shiryaev--Roberts a Headstart}
\titlerunning{Optimal Design of the Shiryaev--Roberts Chart}

\author{Aleksey S. Polunchenko}
\authorrunning{Polunchenko}

\institute{A.S. Polunchenko \at Department of Mathematical Sciences, State University of New York at Binghamton, Binghamton, New York 13902--6000, USA \email{aleksey@binghamton.edu}}

\maketitle
\abstract{We offer a numerical study of the effect of headstarting on the performance of a Shiryaev--Roberts (SR) chart set up to control the mean of a normal process. The study is a natural extension of that previously carried out by Lucas \& Crosier~\citeyearpar{Lucas+Crosier:T1982} for the CUSUM scheme. The Fast Initial Response (FIR) feature exhibited by a headstarted CUSUM turns out to be also characteristic of an SR chart (re-)started off a positive initial score. However, our main result is the observation that a FIR SR with a carefully designed {\em optimal} headstart is not just faster to react to an initial out-of-control situation, it is nearly {\em the} fastest {\em uniformly}, i.e., assuming the process under surveillance is equally likely to go out of control effective any sample number. The performance improvement is the greater, the fainter the change. We explain the optimization strategy, and tabulate the optimal initial score, control limit, and the corresponding ``worst possible'' out-of-control Average Run Length (ARL), considering mean-shifts of diverse magnitudes and a wide range of levels of the in-control ARL.
}
\abstract*{We offer a numerical study of the effect of headstarting on the performance of a Shiryaev--Roberts (SR) chart set up to control the mean of a normal process. The study is a natural extension of that previously carried out by Lucas and Crosier for the CUSUM scheme in their seminal 1982 paper published in Technometrics. The Fast Initial Response (FIR) feature exhibited by a headstarted CUSUM turns out to be also characteristic of an SR chart (re-)started off a nonzero initial score. However, our main result is the observation that a FIR SR with a carefully designed {\em optimal} headstart is not just faster to react to an initial out-of-control situation, it is nearly {\em the} fastest {\em uniformly}, i.e., assuming the process under surveillance is equally likely to go out of control effective any sample number. The performance improvement is the greater, the fainter the change. We explain the optimization strategy, and tabulate the optimal initial score, control limit, and the corresponding ``worst possible'' out-of-control Average Run Length (ARL), considering mean-shifts of diverse magnitudes and a wide range of levels of the in-control ARL.
}

\section{Introduction}\label{c12:sec:intro}
The general theme of this work is the optimal design of the Shiryaev--Roberts (SR) chart originally proposed by Shiryaev~\citeyearpar{Shiryaev:SMD1961,Shiryaev:TPA1963} and Roberts~\citeyearpar{Roberts:T1966}, and later generalized by~\cite{Moustakides+etal:SS2011}. Set up to detect a possible change in the baseline mean of a series of independent samples $X_1,X_2,\ldots$ drawn from a normal unit-variance population at regular time intervals, the classical SR chart involves sequential evaluation of the SR statistic $\{R_n\}_{n\ge0}$ using the recurrence $R_n=(1+R_{n-1})\exp\{S_n\}$, $n=1,2,\ldots$, with $R_0=0$, and where the quantity
\begin{equation}\label{eq:SR-score-func-def}
S_n
\triangleq
\mu\left(X_n-\dfrac{\mu}{2}\right)
\end{equation}
is a numerical score that captures the severity of the deviation of the $n$-th sample point $X_n$ from the target mean-value in either direction; the score function $S_n$ assumes that the intended (target) mean-value of the data is zero, but it is anticipated to change abruptly and permanently to a known off-target value $\mu\neq0$. The $n$-th observation $X_n$ might represent a single reading or the average of a batch of observations from a designated routine sampling plan. The chart triggers an alarm at the first stage, $\mathcal{S}_A$, such that $R_{\mathcal{S}_A}\ge A$, where $A>0$ is a control limit (detection threshold) set in advance in accordance with the desired level of the false alarm risk; more formally, $\mathcal{S}_A\triangleq\min\{n\ge1\colon R_n\ge A\}$, where $A>0$ is given. Hence the process $\{X_n\}_{n\ge1}$ is considered to be in control until stage $\mathcal{S}_A$. The random variable, $\mathcal{S}_A$, referred to as the run length, is the stage at which sampling stops and appropriate action is taken. A brief account of the history of the SR chart was recently offered by Pollak~\citeyearpar{Pollak:IWSM2009}. For an up-to-date summary of the classical as well as generalized SR charts' optimality properties, see, e.g.,~\cite{Polunchenko+Tartakovsky:MCAP2012}.

Though nowhere nearly as known and as widespread as Page's~\citeyearpar{Page:B1954} celebrated CUSUM ``inspection scheme'', the SR chart did receive some attention in the applied literature. One of the earliest investigations of the chart's characteristics is due to~\cite{Roberts:T1966}, who offered a performance comparison of the chart against a host of other statistical process control procedures, including the CUSUM scheme and the EWMA chart \citep[also introduced by][]{Roberts:T1959}. A similar type of SR-vs-CUSUM comparison (but with respect to a different criterion and for a different data model) was also later performed by Mevorach \& Pollak~\citeyearpar{Mevorach+Pollak:AJMMS1991}. See also, e.g.,~\cite{Tartakovsky+Ivanova:PIT1992},~\cite{Tartakovsky+etal:IWSM2009}, and~\cite{Moustakides+etal:CommStat2009}. Certain data-analytic advantages of the chart over the CUSUM scheme were pointed out by Kenett \& Pollak~\citeyearpar{Kenett+Pollak:JAP1996}. Kenett \& Pollak~\citeyearpar{Kenett+Pollak:IEEE-TR1986} provided an example of an application of the SR chart in the area of software reliability.

In the (more theoretical) area of quickest change-point detection, the SR chart received far more attention. To a large extent this is due to the fundamental work of Shiryaev~\citeyearpar{Shiryaev:SMD1961,Shiryaev:TPA1963} who proved that the chart solves a particular Bayesian version of the quickest change-point detection problem; see also~\cite{Girshick+Rubin:AMS1952}. The chart then remained unnoticed until recently Pollak \& Tartakovsky~\citeyearpar{Pollak+Tartakovsky:SS2009} and Shiryaev \& Zryumov~\citeyearpar{Shiryaev+Zryumov:Khabanov2010} discovered that it solves yet another so-called {\em multi-cyclic} or {\em generalized} Bayesian version of the quickest change-point detection problem; the multi-cyclic setup is instrumental in such applications as cybersecurity~\citep[see, e.g.,][]{Tartakovsky+etal:JSTSP2013}, financial monitoring~\citep[see, e.g.,][]{Pepelyshev+Polunchenko:SII2016}, and economic design of control charts. This brought the SR chart back into the spotlight. Polunchenko et al.~\citeyearpar{Polunchenko+etal:CommStat2016} performed a robustness analysis of the SR chart's multi-cyclic capabilities when the post-change distribution involves a misspecified parameter. Moustakides et al.~\citeyearpar{Moustakides+etal:SS2011} observed that by starting the SR statistic $\{R_n\}_{n\ge0}$ off a positive initial value, i.e., setting $R_0=r>0$, the SR chart can be made nearly the best \citep[in the minimax sense of][]{Pollak:AS1985}. Roughly, this means the SR chart is almost the fastest to react to a change in the observations' distribution when the corresponding unknown change-point is equally likely to be any point in time; see Section~\ref{c12:sec:preliminaries} for a formal definition. As a matter of fact Polunchenko \& Tartakovsky~\citeyearpar{Polunchenko+Tartakovsky:AS2010} and Tartakovsky \& Polunchenko~\citeyearpar{Tartakovsky+Polunchenko:IWAP2010} demonstrated that in two specific change-point scenarios the SR chart with a carefully designed headstart is {\em the} fastest \citep[in the sense of][]{Pollak:AS1985}. This result was then extended by Tartakovsky et al.~\citeyearpar{Tartakovsky+etal:TPA2012} who proved that the SR chart whose headstart is selected in a specific fashion is almost the best one can do \citep[again, in the sense of][]{Pollak:AS1985} asymptotically, as the false alarm risk tends to zero, in a general change-point scenario.

In spite of the aforementioned strong theoretically established optimality properties of the SR chart, and the fact that no such properties are exhibited by either the CUSUM scheme or the EWMA chart, applications of the SR chart in quality control remain essentially nonexistent. In part, this may be due to the lack of existing resources with pre-computed, for a variety of cases, optimal headstart and control limit values. To the best of our knowledge, the work of Tartakovsky et al.~\citeyearpar{Tartakovsky+etal:IWSM2009} and that of Polunchenko \& Sokolov~\citeyearpar{Polunchenko+Sokolov:EnT2014} have heretofore been the only sources with such data (computed assuming the observations are exponential). This work's goal is to optimize the SR chart for yet another model, namely, the standard Gaussian model widely used in the quality control literature as a testbed for charts' performance analysis. The specific optimization strategy is presented in Section~\ref{c12:sec:preliminaries}. The optimization itself is carried out in Section~\ref{c12:sec:main-result} using the numerical framework developed by Moustakides et al.~\citeyearpar{Moustakides+etal:SS2011} and then improved upon by Polunchenko et al.~\citeyearpar{Polunchenko+etal:SA2014,Polunchenko+etal:ASMBI2014}. The obtained optimal headstart and control limit values are reported in Section~\ref{c12:sec:main-result} as well. Conclusions follow in Section~\ref{c12:sec:conclusion}.

\section{The Shiryaev--Roberts Chart, Its Properties and Optimization}\label{c12:sec:preliminaries}
To control the mean of a standard Gaussian process, the headstarted tweak of the classical SR chart proposed by Moustakides et al.~\citeyearpar{Moustakides+etal:SS2011} operates by sequentially updating the statistic $\{R_n^r\}_{n\ge0}$ via the recurrence
\begin{equation}\label{eq:Rn-SR-r-def}
R_{n}^{r}
=
(1+R_{n-1}^{r})\exp\{S_n\},\;n=1,2,\ldots\;\text{with}\; R_{0}^{r}=r\ge0,
\end{equation}
where $S_n$ is the score function defined in~\eqref{eq:SR-score-func-def}; the initial score $R_0^r=r\ge0$ is a design parameter also referred to as the headstart, which is the original terminology of Lucas \& Crosier~\citeyearpar{Lucas+Crosier:T1982} who suggested to headstart the CUSUM scheme. The corresponding run length is as follows:
\begin{equation}\label{eq:T-SR-r-def}
\mathcal{S}_{A}^{r}
\triangleq
\min\{n\ge1\colon R_{n}^{r}\ge A\},
\end{equation}
where $A>0$ is the control limit (detection threshold) selected in advance so as to keep the chart's false alarm characteristics tolerably low. Note that if $r=0$ then the chart is the classical SR chart (with no headstart) of Shiryaev~\citeyearpar{Shiryaev:SMD1961,Shiryaev:TPA1963} and Roberts~\citeyearpar{Roberts:T1966}. For this reason Tartakovsky et al.~\citeyearpar{Tartakovsky+etal:TPA2012} coined the term ``{\em Generalized} SR chart'' (or the GSR chart for short) to refer to the headstarted SR chart defined by~\eqref{eq:Rn-SR-r-def} and~\eqref{eq:T-SR-r-def}. It is also worth reiterating that the score function~\eqref{eq:SR-score-func-def}---and hence also the statistic~\eqref{eq:Rn-SR-r-def}---are indifferent to the direction of the mean shift, i.e., the sign of $\mu\neq0$ is irrelevant.

It has been the custom in the quality control literature to assess the operating characteristics of a control chart, with run length $\T$, by means of only two indices: the in-control Average Run Length (ARL) and the out-of-control ARL. In this work, we shall adapt the (more exhaustive) approach used in the quickest change-point detection literature. Let $\Pr_k$ ($\EV_k$) denote the probability measure (expectation) induced by the data $\{X_n\}_{n\ge1}$ assuming the change-point is at time moment $k=0,1,2,\ldots,\infty$, i.e., assuming the process $\{X_n\}_{n\ge1}$ is in-control until sample number $k$ inclusive, and is out-of-control starting from sample number $k+1$ onward. The notation $k=0$ ($k=\infty$) is to be understood as the case when the process under surveillance is out of control {\it ab initio} (never, respectively).

The in-control characteristics of a control chart $\T$ are usually gauged by virtue of the Average Run Length (ARL) to false alarm $\ARL(\T)\triangleq\EV_\infty[\T]$ which is the average number of samples taken by the chart before an {\em erroneous} out-of-control signal is given; this is precisely what is known in the quality control literature as the {\em in-control} ARL. It is apparent that the higher the ARL to false alarm, the lower the level of the false alarm risk. For the GSR chart, the general inequality $\ARL(\mathcal{S}_A^r)\ge A-r$ can be used to design $A>0$ and $r\in[0,A]$ so as to have $\ARL(\mathcal{S}_{A}^{r})$ no lower than a desired margin $\gamma>1$. It is of note that this inequality holds in general, whatever the statistical structure of the observations be. A more accurate result is the asymptotic (as $A\to+\infty$) approximation $\ARL(\mathcal{S}_A^r)\approx A/\xi-r$, which is actually known to be quite accurate even if $A>0$ is not high; see, e.g.,~\cite[Theorem~1]{Pollak:AS1987} or~\cite{Tartakovsky+etal:TPA2012}. Here $\xi$ denotes the so-called ``limiting average exponential overshoot''---a model-dependent constant (taking values between 0 and 1) computable using nonlinear renewal-theoretic methods; see, e.g.,~\cite{Woodroofe:Book1982}. For the Gaussian model considered in this work it follows, e.g., from~\cite[Example~3.1,~pp.~32--33]{Woodroofe:Book1982}, that the following formula can be used:
\begin{equation}\label{eq:xi-formula}
\xi
=
\dfrac{2}{\mu^2}\exp\left\{-2\sum_{m=1}^{\infty}\dfrac{1}{m}\Phi\left(-\dfrac{\mu}{2}\sqrt{m}\right)\right\},
\end{equation}
where
\begin{equation*}
\Phi(x)
\triangleq
\dfrac{1}{\sqrt{2\pi}}\int_{-\infty}^{x}e^{-\tfrac{t^2}{2}}\,dt
\end{equation*}
is the standard Gaussian cumulative distribution function. Note from the foregoing formula that $\xi$ is an even function of $\mu\neq0$. The formula was put to use by Woodroofe~\citeyearpar{Woodroofe:Book1982} who computed $\xi$ for various values of $\mu>0$; see~\cite[Table~3.1,~p.~33]{Woodroofe:Book1982} for the obtained results.

To quantify the capabilities of a control chart $\T$ when the process is no longer in control, Pollak~\citeyearpar{Pollak:AS1985} suggested to use the ``worst-case'' (Supremum) Average Detection Delay (SADD), conditional on no false alarm having been sounded. Formally,
\begin{equation*}
\SADD(\T)
\triangleq
\max_{0\le k<\infty}\ADD_{k}(\T),
\end{equation*}
where $\ADD_{k}(\T)\triangleq\EV_k[\T-k|\T>k]$, $k=0,1,2,\ldots$. Incidentally, the limiting ADD value $\lim_{k\to\infty}\ADD_k(\T)$ is known in the quality control literature as the {\em steady-state} ARL.

Pollak's~\citeyearpar{Pollak:AS1985} criterion has a simple interpretation: for any fixed but finite $k=0,1,2,\ldots$, the condition $\T>k$ guarantees that it is an actual detection (not a false alarm), so that each $\ADD_k(\T)$ is the average number of samples it takes the chart past the change-point $k$ to realize the process is not in control anymore, and because $k$ is unknown, it is reasonable to assume it equally likely to be any number ($0,1,2,\ldots$) and consider the worst possible case, i.e., take the maximal of the $\ADD_k(\T)$'s. For the CUSUM scheme with no headstart and for the classical SR chart (also headstart-free) it can be shown that $k=0$ is when the ADD is the highest, i.e., $\SADD(\T)=\ADD_0(\T)$. As a result, it suffices to restrict attention to just $\ADD_0(\T)$, and it is this quantity that the quality control community calls the out-of-control ARL. However, things are not as simple when the chart has a positive headstart, and it is no longer obvious which of the delays $\ADD_k(\mathcal{S}_A^r)$'s for $k=0,1,2,\ldots$ is the highest. As a matter of fact we shall see in the next section that the ``bump'' of the sequence $\{\ADD_k(\mathcal{S}_A^r)\}_{k\ge0}$ has a highly unpredictable behavior in terms of its location on the time axis.

Let $\Delta(\gamma)\triangleq\{\T\colon\ARL(\T)\ge\gamma\}$ be the class of control charts (identified with a generic run length $\T$) whose ARL to false alarm is at least as high as a desired pre-set level $\gamma>1$. Pollak's~\citeyearpar{Pollak:AS1985} minimax change-point detection problem consists in finding $\T_{\mathrm{opt}}\in\Delta(\gamma)$ such that $\SADD(\T_{\mathrm{opt}})=\min_{\T\in\Delta(\gamma)}\SADD(\T)$ for any given $\gamma>1$. In general, this problem is still an open one, although there has been a continuous effort to solve it. To that end, for at least two specific data models, the answer was shown to be the GSR chart with ``finetuned'' threshold and headstart values; see~\cite{Polunchenko+Tartakovsky:AS2010} and~\cite{Tartakovsky+Polunchenko:IWAP2010}. Moreover, for a general data model, the GSR chart (properly optimized) was also shown (by~\citealt{Tartakovsky+etal:TPA2012}) to solve Pollak's~\citeyearpar{Pollak:AS1985} problem asymptotically as $\gamma\to+\infty$. Specifically, this means that if $A$ and $r$ are selected so that $\ARL(\mathcal{S}_A^r)\ge\gamma$ with $\gamma>1$ given, i.e., $\mathcal{S}_{A}^{r}\in\Delta(\gamma)$, then
\begin{equation}\label{eq:SR-r-SADD-optimality-order3}
\SADD(\mathcal{S}_{A}^{r})-\min_{\T\in\Delta(\gamma)}\SADD(\T)\searrow 0\;\;\text{as}\;\;\gamma\to+\infty,
\end{equation}
provided, however, that $r/A\to 0$ as $A\to+\infty$; see~\cite{Tartakovsky+etal:TPA2012}, who also supply a high-order large-$\gamma$ expansion of $\SADD(\mathcal{S}_{A}^{r})$. The foregoing is a strong optimality property known in the literature on change-point detection as asymptotic minimax optimality of order three, or asymptotic near minimaxity. It is noteworthy that the CUSUM chart, whether headstarted or not, does not have such strong ``nearly-best'' detection capabilities. Moreover, nor does the EWMA chart. Hence, our interest in the GSR chart.  To provide an idea as to the difference made by a positive headstart, we remark that the classical SR chat (with zero headstart) is asymptotically (as $\gamma\to+\infty$) minimax of order two, i.e., the difference $\SADD(\mathcal{S}_{A})-\min_{\T\in\Delta(\gamma)}\SADD(\T)$ goes to a positive constant as $\gamma\to+\infty$. Moreover, since the constant is the higher, the fainter the change, giving an SR chart a positive headstart is especially beneficial when the out-of-control behavior of the process differs from its in-control behavior only slightly.

Yet another strong optimality property of the GSR chart is its exact multi-cyclic or generalized Bayesian optimality. Specifically, Pollak \& Tartakovsky~\citeyearpar{Pollak+Tartakovsky:SS2009} and Shiryaev \& Zryumov~\citeyearpar{Shiryaev+Zryumov:Khabanov2010} proved that the classical SR chart (with no headstart) minimizes the so-called Integral ADD
\begin{equation}\label{eq:IADD-def}
\IADD(\T)
\triangleq
\sum_{k=0}^{\infty}\EV_k[\max\{0,\T-k\}],
\end{equation}
and the so-called Relative IADD (RIADD)
\begin{equation}\label{eq:RIADD-def}
\RIADD(\T)
\triangleq
\IADD(\T)/\ARL(\T)
=
\sum_{k=0}^{\infty}\dfrac{\Pr_{\infty}(\T>k)}{\ARL(\T)}\,\ADD_k(\T),
\end{equation}
both inside the class $\Delta(\gamma)$ defined above, for any $\gamma>1$. The meaning of this result can be explained by analyzing the structure of the definition~\eqref{eq:RIADD-def} of $\RIADD(\T)$. Specifically, on the one hand, the latter can be viewed as being the $k$-average of the delays $\EV_k[\max\{0,\T-k\}]$, $k=0,1,2,\ldots$, assuming that change-point $k$ has an improper uniform distribution on the set $\{0,1,2,\ldots\}$. The improper uniformity of the change-point is a core assumption of the generalized Bayesian change-point detection problem. On the other hand, $\RIADD(\T)$ can also be regarded as the $k$-average of the $\ADD_k(\T)$'s assuming that the probability mass function of $k$ is given by the ratio $\Pr_{\infty}(\T>k)/\ARL(\T)$, $k=0,1,2,\ldots$; note that $\Pr_{k}(\T>k)\equiv\Pr_{\infty}(\T>k)$ for any $k=0,1,2,\ldots$, and that $\ARL(\T)=\sum_{k=0}^{\infty}\Pr_{\infty}(\T>k)$. For yet another, viz. multi-cyclic interpretation, see~\cite{Pollak+Tartakovsky:SS2009}.

The $\RIADD$-optimality of the classical SR chart was generalized in~\cite[Lemma~1]{Polunchenko+Tartakovsky:AS2010} where it was shown that the GSR chart, whose control limit $A>0$ and headstart $r\ge0$ are such that $\ARL(\mathcal{S}_{A}^{r})\ge\gamma$ for a given $\gamma>1$, minimizes the so-called Stationary ADD (STADD)
\begin{equation}\label{eq:STADD-gen-def}
\STADD(\T)
\triangleq
\left(r\ADD_0(\T)+\IADD(\T)\right)\left/\left(\ARL(\T)+r\right)\right.
\end{equation}
inside class $\Delta(\gamma)$, for any $\gamma>1$; recall that $\IADD(\T)$ is as in~\eqref{eq:IADD-def}. Formally, for any $\gamma>1$, and any $A>0$ and $r\ge0$, it holds true that $\STADD(\mathcal{S}_{A}^{r})=\min_{\T\in\Delta(\gamma)}\STADD(\T)$, provided that $\ARL(\mathcal{S}_{A}^{r})\ge\gamma$ is satisfied. Also, observe that $\STADD(\mathcal{S}_{A}^{r})$ reduces to $\RIADD(\mathcal{S}_{A}^{r})$ when $r=0$. It is also of note that $\STADD(\T)$ is not the same as the limit $\lim_{k\to\infty}\ADD_k(\T)$.

An important ``by-product'' of~\cite[Lemma~1]{Polunchenko+Tartakovsky:AS2010} is that the quantity $\STADD(\mathcal{S}_{A}^{r})$ turns out to also provide a universal lowerbound on the unknown value of $\min_{\T\in\Delta(\gamma)}\SADD(\T)$, and this lowerbound is valid for any $\gamma>1$ and $r\ge0$ such that $\mathcal{S}_{A}^{r}\in\Delta(\gamma)$; see~\cite[Theorem~1]{Polunchenko+Tartakovsky:AS2010}. Specifically, introducing $\underline{\SADD}(\mathcal{S}_{A}^{r})\equiv\STADD(\mathcal{S}_{A}^{r})$, the following double inequality holds:
\begin{equation}\label{eq:SADD-double-inequality}
\underline{\SADD}(\mathcal{S}_{A}^{r})
\le
\min_{\T\in\Delta(\gamma)}\SADD(\T)
\le
\SADD(\mathcal{S}_{A}^{r}),
\end{equation}
for any $A>0$ and $r\ge0$ such that $\ARL(\mathcal{S}_A^r)\ge\gamma$, and any given $\gamma>1$; cf.~\cite[Inequality~(2.12),~p.~579]{Moustakides+etal:SS2011}.

A few important comments are now in order:
\begin{enumerate}
    \setlength{\itemsep}{3pt}
    \item On the one hand, the double inequality~\eqref{eq:SADD-double-inequality}, namely, its left part, implies that the lowerbound $\underline{\SADD}(\mathcal{S}_{A}^{r})\equiv\STADD(\mathcal{S}_{A}^{r})$, where $\STADD(\T)$ is defined in~\eqref{eq:STADD-gen-def}, can be used as a benchmark to get an idea as to how much room there is for improvement in the way of SADD for a chart of interest. Should it so happen that the SADD of the chart of interest with the ARL to false alarm level set to $\gamma>1$ is only a tiny bit greater than $\underline{\SADD}(\mathcal{S}_{A}^{r})$ assuming $\ARL(\mathcal{S}_{A}^{r})=\gamma>1$, then the chart is almost minimax optimal in the sense of Pollak~\citeyearpar{Pollak:AS1985}.
    \item On the other hand, the double inequality~\eqref{eq:SADD-double-inequality} also suggests the following optimization strategy for the GSR chart: for a given $\gamma>1$, pick the chart's detection threshold $A>0$ and headstart $r\ge0$ in such a way so as to make the difference $\SADD(\mathcal{S}_{A}^{r})-\underline{\SADD}(\mathcal{S}_{A}^{r})$ as close to zero as is possible without violating the inequality $\ARL(\mathcal{S}_{A}^{r})\ge\gamma$. More formally, the optimal detection threshold $A^{*}$ and headstart $r^{*}$ values are to be selected as follows:
\begin{equation}\label{eq:opt-r-A-choice}
(r^{*},A^{*})
=
\argmin_{r,A\ge 0}\left\{\SADD(\mathcal{S}_{A}^{r})-\underline{\SADD}(\mathcal{S}_{A}^{r})\right\},
\;\text{but}\;
\ARL(\mathcal{S}_{A}^{r})=\gamma,
\end{equation}
where $\gamma>1$ is given; it goes without saying that both $A^{*}$ and $r^{*}$ are functions of $\gamma>1$. The foregoing optimization strategy is originally due to~\cite{Moustakides+etal:SS2011}, and, in this work, we shall adapt it as well.
    \item As we shall demonstrate in the next section, if the GSR chart's detection threshold $A$ and initial score $r$ are set to $A^{*}$ and $r^{*}$, respectively, where $A^{*}$ and $r^{*}$ are as in~\eqref{eq:opt-r-A-choice} with $\gamma>1$ given, then, conditional on $\ARL(\mathcal{S}_{A}^{r})=\gamma$, the difference $\SADD(\mathcal{S}_{A}^{r})-\underline{\SADD}(\mathcal{S}_{A}^{r})$ is nearly zero, even if $\gamma>1$ is on the order of hundreds. Therefore, the GSR chart's third-order asymptotic optimality~\eqref{eq:SR-r-SADD-optimality-order3} does not necessarily require $\gamma$ to be large.
\end{enumerate}

The constrained optimization problem~\eqref{eq:opt-r-A-choice} can be solved numerically, e.g., with the aid of the numerical method proposed by Moustakides et al.~\citeyearpar{Moustakides+etal:SS2011} and subsequently improved upon by Polunchenko et al.~\citeyearpar{Polunchenko+etal:SA2014,Polunchenko+etal:ASMBI2014}. This is precisely the object of the next section.

\section{Experimental Results}\label{c12:sec:main-result}
The plan now is to employ the numerical framework of Moustakides et al.~\citeyearpar{Moustakides+etal:SS2011} and its improved version due to Polunchenko et al.~\citeyearpar{Polunchenko+etal:SA2014,Polunchenko+etal:ASMBI2014}, and analyze the performance of the GSR chart given by~\eqref{eq:Rn-SR-r-def} and~\eqref{eq:T-SR-r-def} under different parameter settings, including (and especially) the optimal choice given by the solution of the constrained optimization problem~\eqref{eq:opt-r-A-choice}.

\begin{sidewaysfigure}[p]
    \centering
    \subfloat[$\mu=0.2$.]{\label{fig:ARL_vs_r_A__mu02}
        \includegraphics[width=0.5\textwidth]{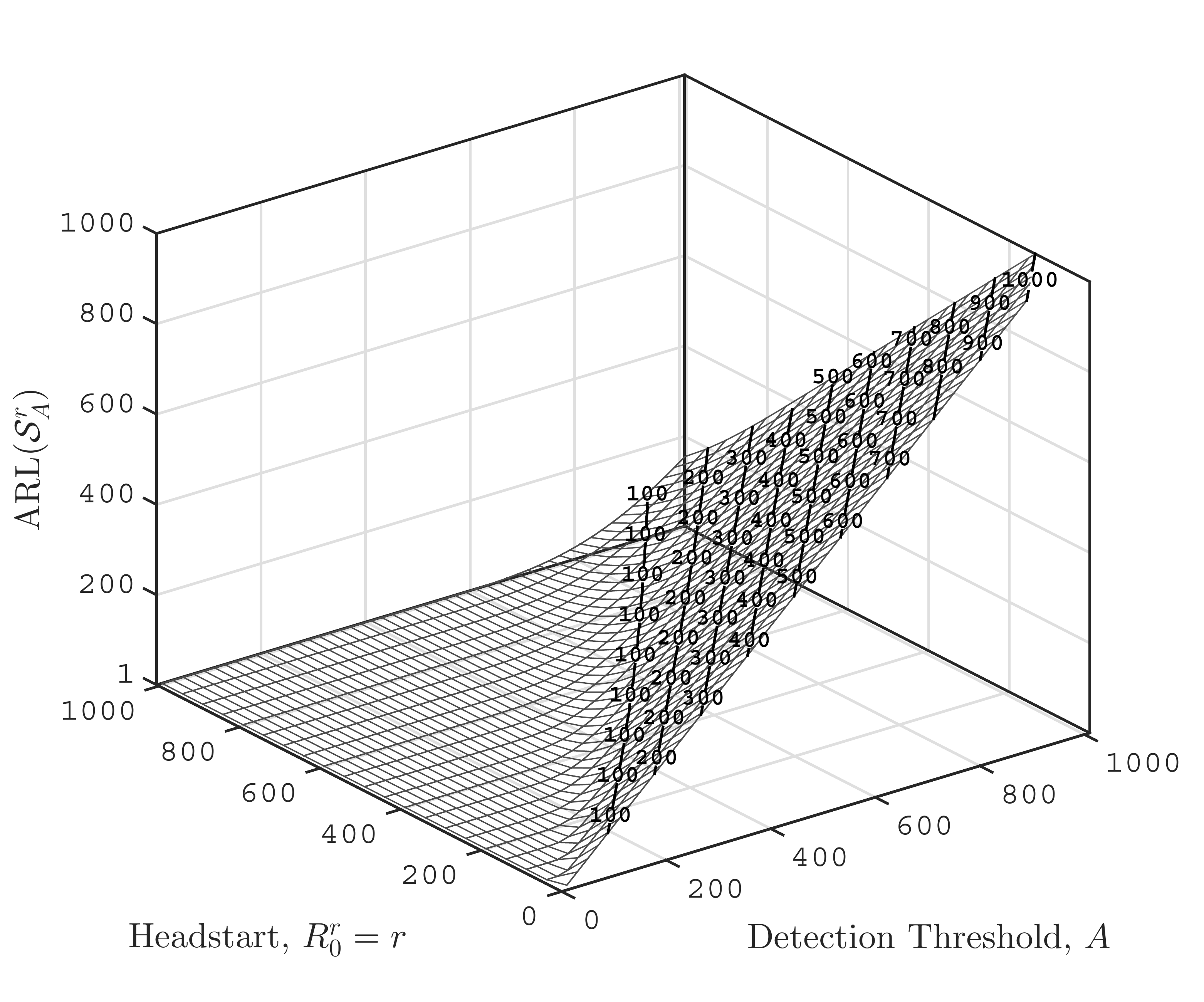}
    } %
    \subfloat[$\mu=0.5$.]{\label{fig:ARL_vs_r_A__mu05}
        \includegraphics[width=0.5\textwidth]{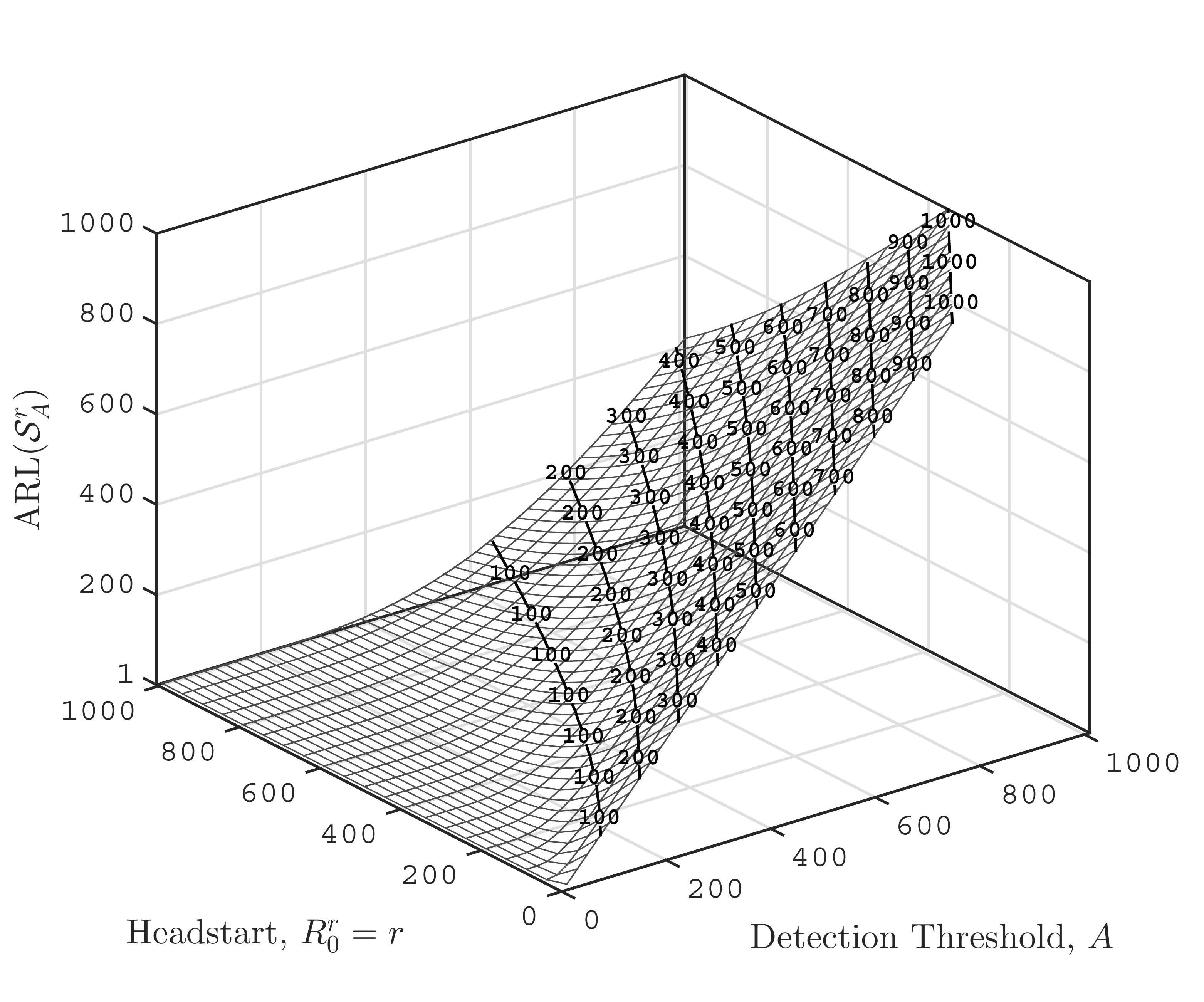}
    } %
    \caption{$\ARL(\mathcal{S}_{A}^{r})$ as a function of the headstart $R_0^r=r\ge0$ and the detection threshold $A>0$ for $\mu=\{0.2,0.5\}$.}
    \label{fig:ARL_vs_r_A}
\end{sidewaysfigure}

We begin with an examination of the level of the ARL to false alarm, i.e., $\ARL(\mathcal{S}_{A}^{r})$, treated as a function of the headstart $r\ge0$, the detection threshold $A>0$,  and the magnitude of the change in the mean $\mu\neq0$. With regard to the latter, for lack of space, let us consider only two cases: $\mu=0.2$ and $\mu=0.5$. The former may be considered a faint change, while the latter is a moderate change. Figures~\ref{fig:ARL_vs_r_A} depict $\ARL(\mathcal{S}_{A}^{r})$ as a function of $r\in[0,A]$ and $A\in[0,1\,000]$. Specifically, Figure~\ref{fig:ARL_vs_r_A__mu02} is for $\mu=0.2$ and Figure~\ref{fig:ARL_vs_r_A__mu05} is for $\mu=0.5$. As can be seen from either figure, the bivariate function $\ARL(\mathcal{S}_{A}^{r})$ is almost linear in $A$ (with $r$ fixed) as well as in $r$ (with $A$ fixed). This is in perfect agreement with the aforementioned fact that $\ARL(\mathcal{S}_{A}^{r})\approx A/\xi-r$ where $\xi$ is given by~\eqref{eq:xi-formula}. Since, according to~\cite[Table~3.1,~p.~33]{Woodroofe:Book1982}, the value of $\xi$ for $\mu=0.2$ is roughly $0.89004$ versus approximately $0.74762$ for $\mu=0.5$, the sensitivity of the ARL to false alarm level to the detection threshold is higher, the stronger the change. Figures~\ref{fig:ARL_vs_r_A} also include contours (shown as bold dark curves) corresponding the different fixed levels $\gamma>1$ of the ARL to false alarm. Specifically, each of the contours is the solution set $(r,A)$ of the equation $\ARL(\mathcal{S}_{A}^{r})=\gamma$ for the appropriate value of $\gamma=\{100,200,\ldots,900,1\,000\}$. These contours are important because the process of optimization of the GSR chart begins with picking a value for $\gamma>1$, and then, with $\gamma>1$ set and fixed, restricting attention to only those values of $A>0$ and $r\ge0$ for which the constraint $\ARL(\mathcal{S}_{A}^{r})=\gamma$ is satisfied. Due to space limitations, in this work we shall consider only three values of $\gamma$, namely, $\gamma=\{100,500,1\,000\}$.

\begin{sidewaysfigure}[p]
    \centering
    \subfloat[$\gamma=100$.]{\label{fig:ADDk_vs_r_k__3D_mu02_ARL100}
        \includegraphics[width=0.33\textwidth]{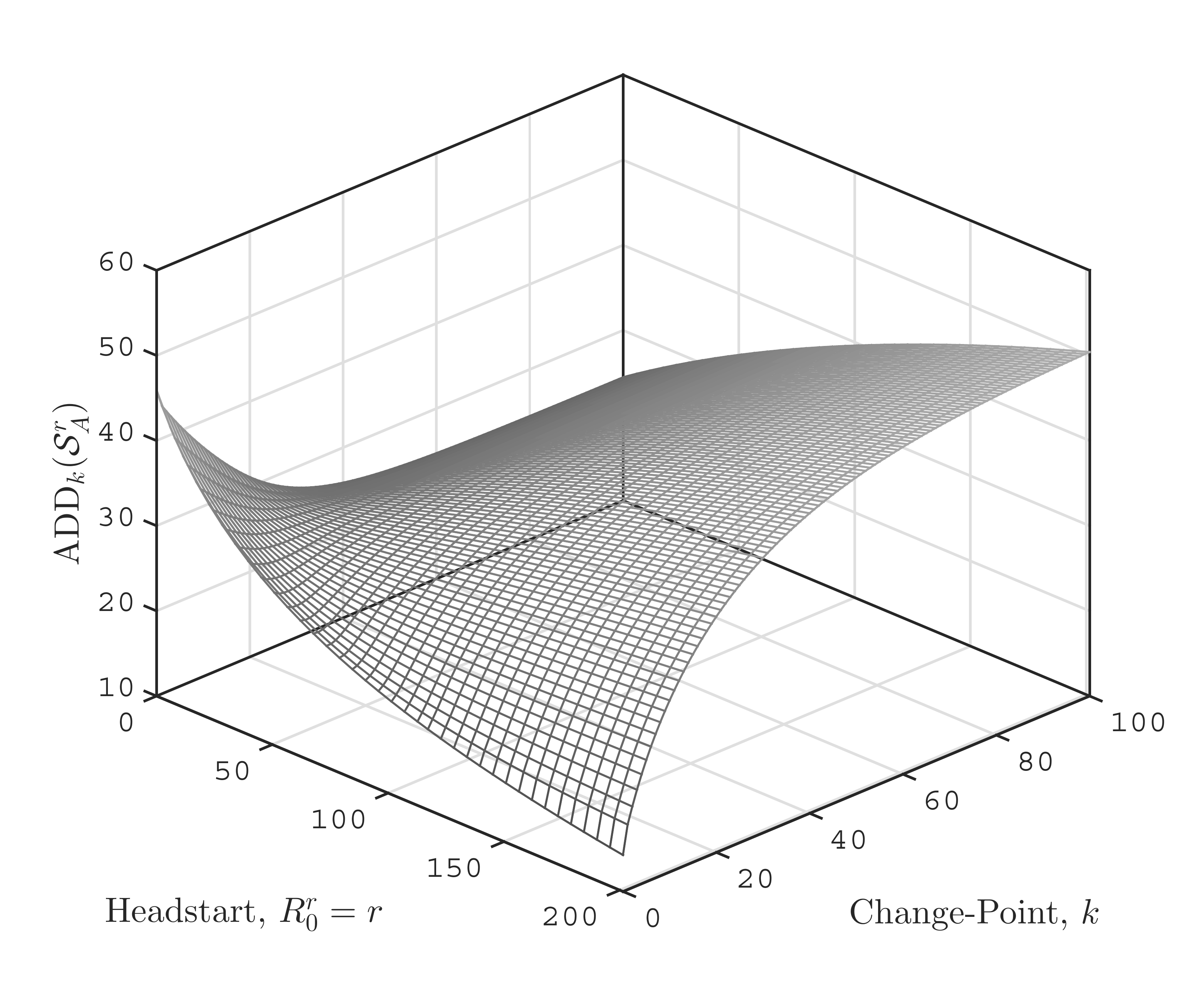}
    } %
    \subfloat[$\gamma=500$.]{\label{fig:ADDk_vs_r_k__3D_mu02_ARL500}
        \includegraphics[width=0.33\textwidth]{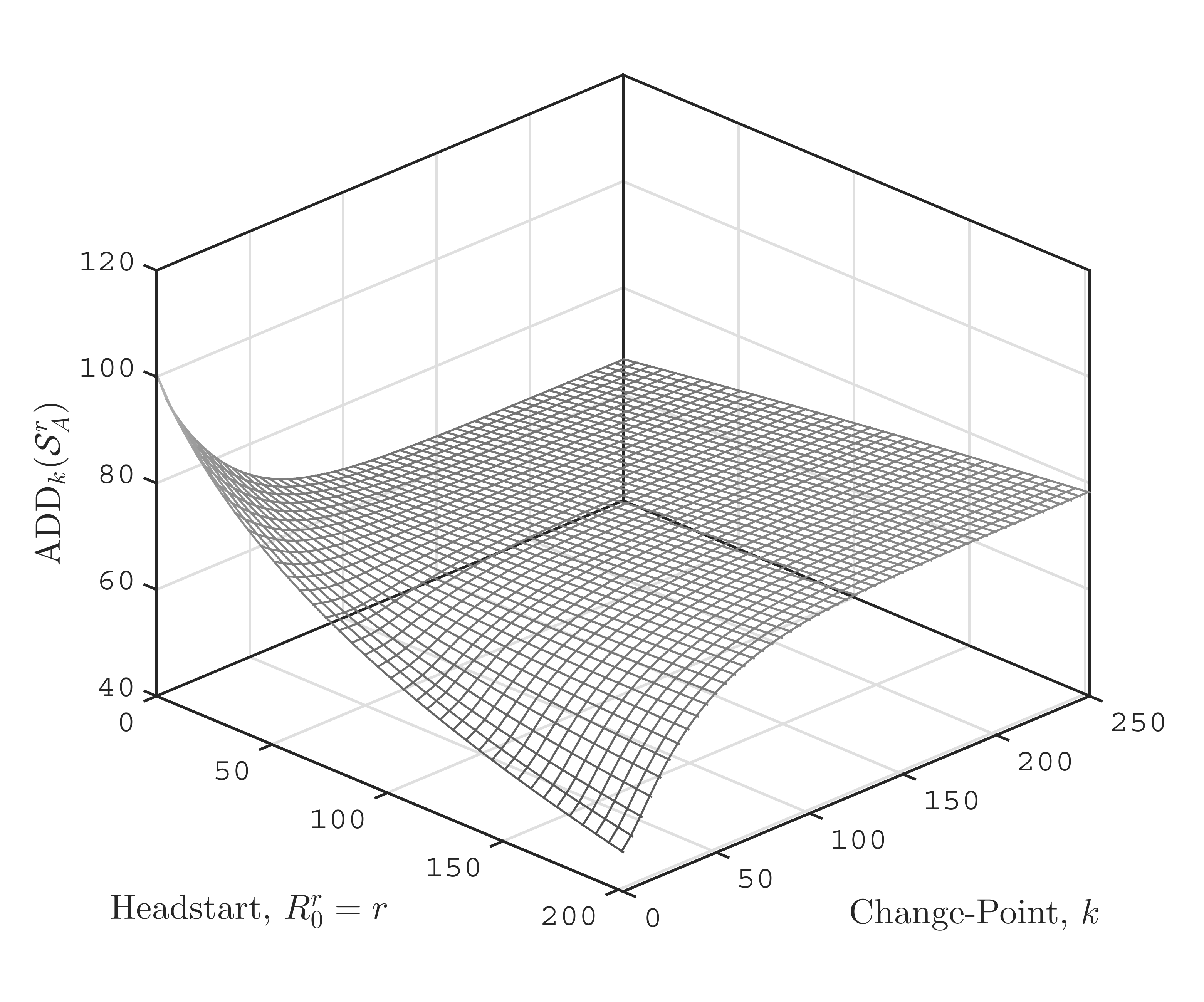}
    } %
    \subfloat[$\gamma=1\,000$.]{\label{fig:ADDk_vs_r_k__3D_mu02_ARL1000}
        \includegraphics[width=0.33\textwidth]{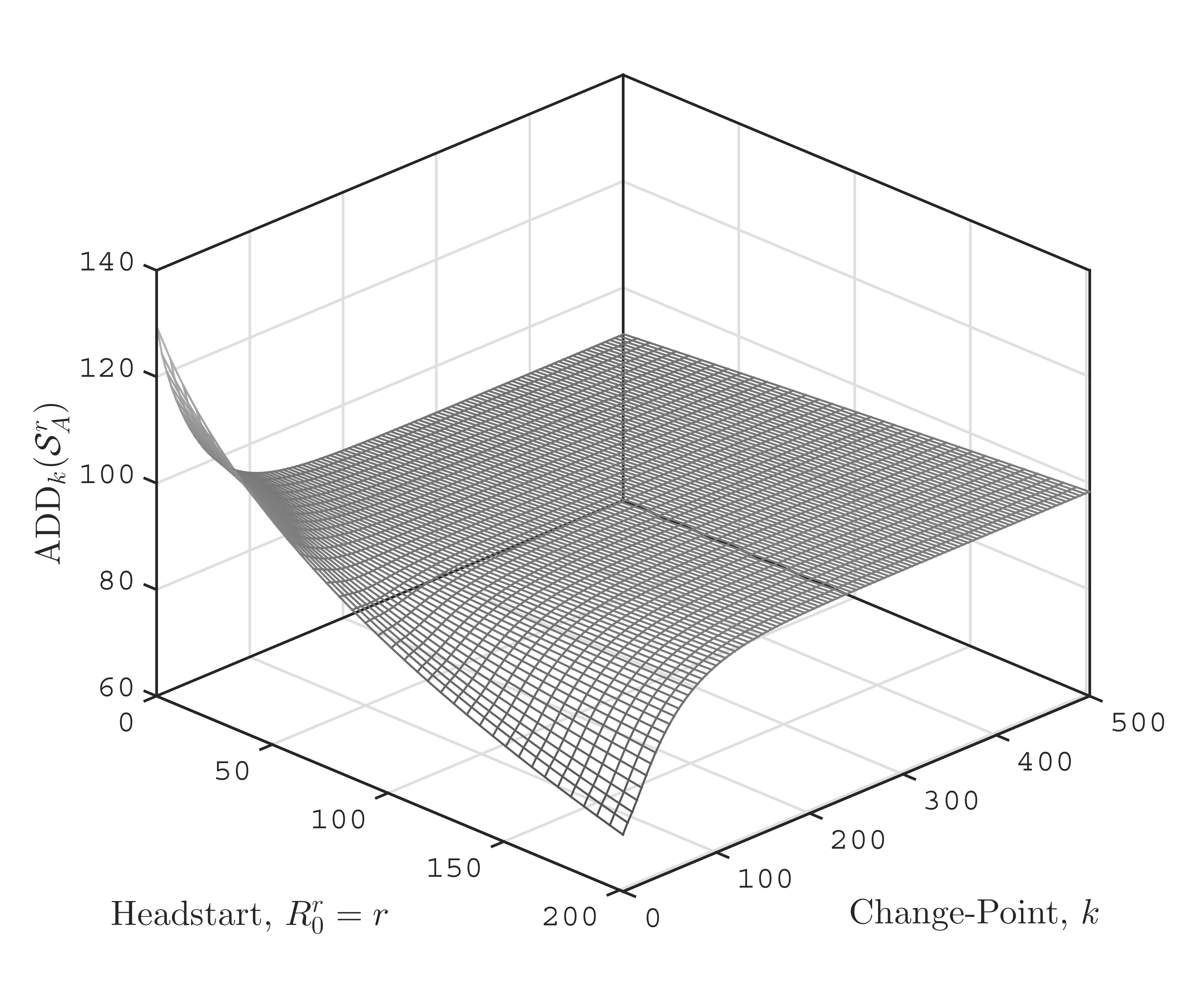}
    } %
    \caption{$\ADD_k(\mathcal{S}_{A}^{r})$ as a function of the headstart $R_0^r=r\ge0$, the change-point $k=0,1,\ldots$, and the ARL to false alarm level $\ARL(\mathcal{S}_{A}^{r})=\gamma>1$ for $\mu=0.2$.}
    \label{fig:ADDk_vs_r_k_ARL__3D_mu02}
\end{sidewaysfigure}

\begin{sidewaysfigure}[p]
    \centering
    \subfloat[$\gamma=100$.]{\label{fig:ADDk_vs_r_k__3D_mu05_ARL100}
        \includegraphics[width=0.33\textwidth]{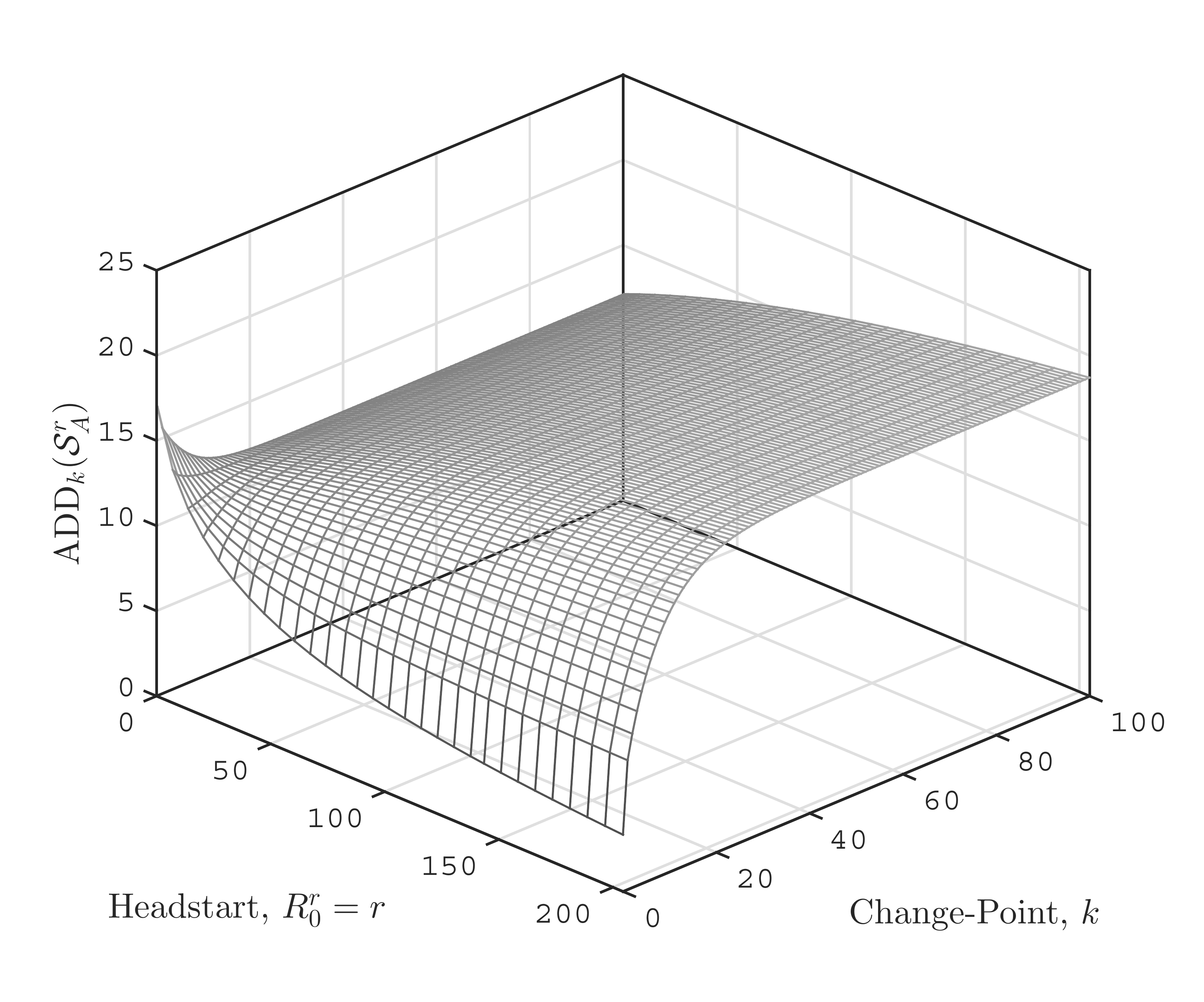}
    } %
    \subfloat[$\gamma=500$.]{\label{fig:ADDk_vs_r_k__3D_mu05_ARL500}
        \includegraphics[width=0.33\textwidth]{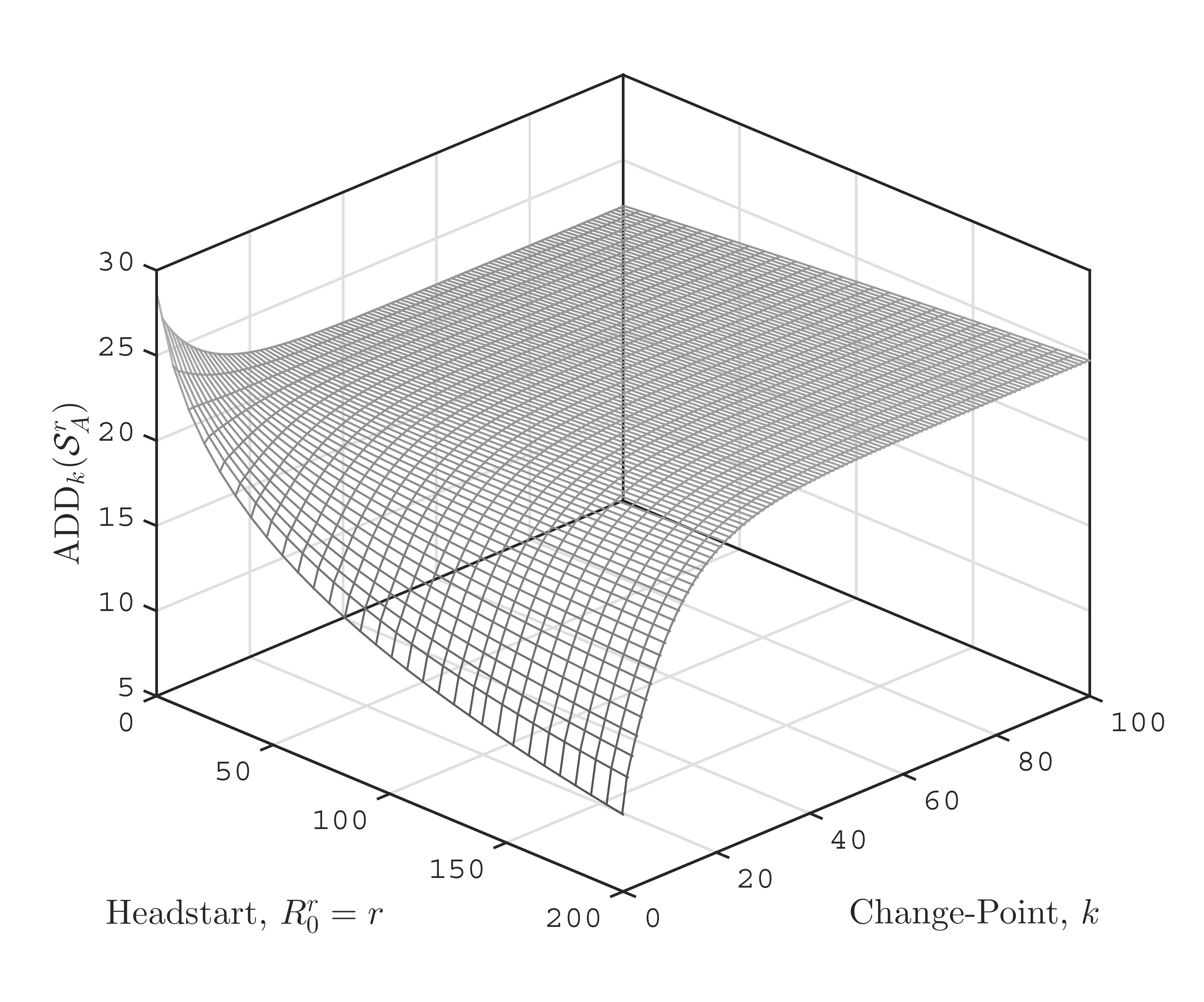}
    } %
    \subfloat[$\gamma=1\,000$.]{\label{fig:ADDk_vs_r_k__3D_mu05_ARL1000}
        \includegraphics[width=0.33\textwidth]{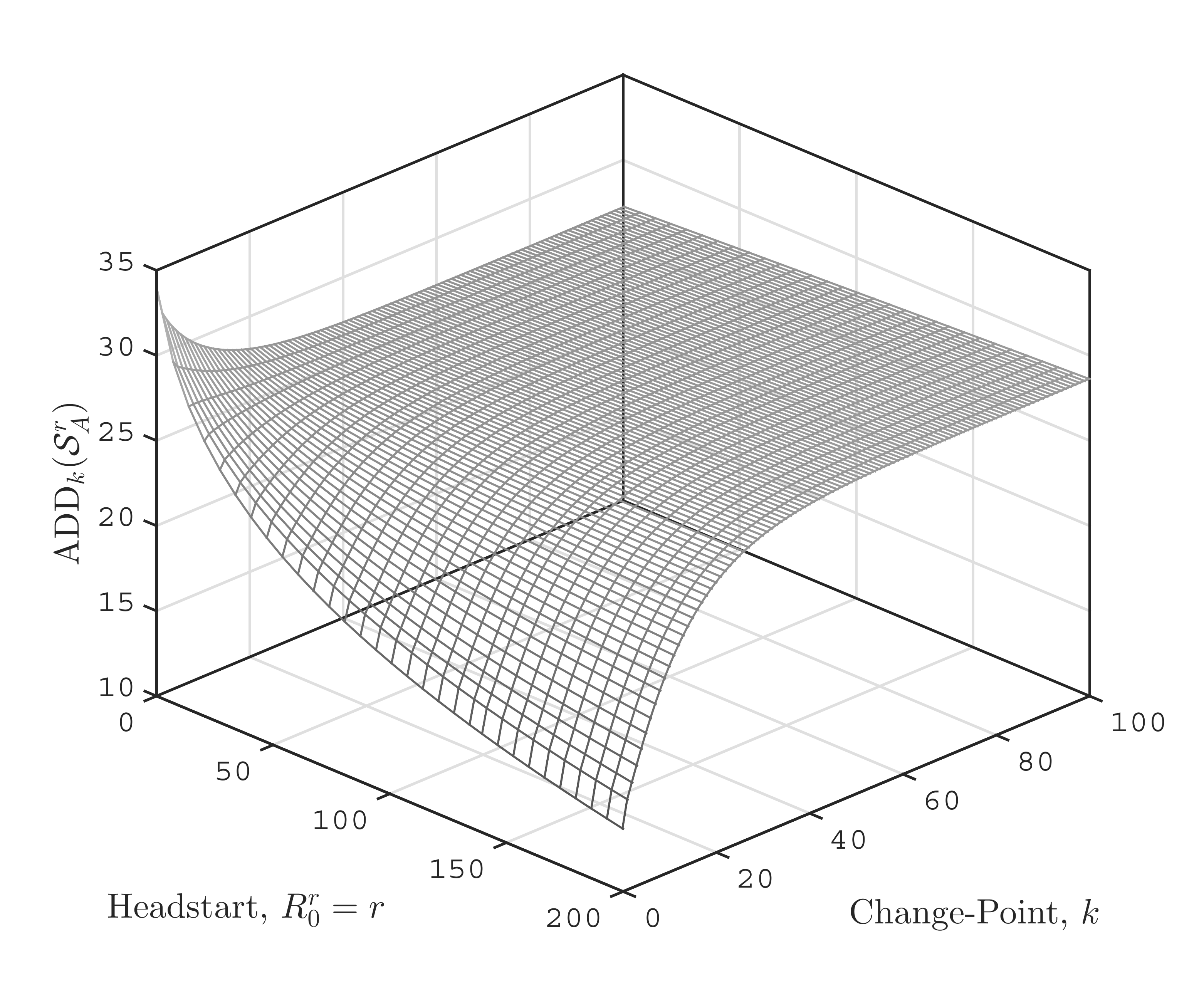}
    } %
    \caption{$\ADD_k(\mathcal{S}_{A}^{r})$ as a function of the headstart $R_0^r=r\ge0$, the change-point $k=0,1,\ldots$, and the ARL to false alarm level $\ARL(\mathcal{S}_{A}^{r})=\gamma>1$ for $\mu=0.5$.}
    \label{fig:ADDk_vs_r_k_ARL__3D_mu05}
\end{sidewaysfigure}

Let us next look at Figures~\ref{fig:ADDk_vs_r_k_ARL__3D_mu02} and~\ref{fig:ADDk_vs_r_k_ARL__3D_mu05} which show $\ADD_k(\mathcal{S}_{A}^{r})$ as a function of $r\ge0$ and $k=0,1,2,\ldots$ under the constraint $\ARL(\mathcal{S}_{A}^{r})=\gamma$ with $\gamma=\{100,500,1\,000\}$. Specifically, Figures~\ref{fig:ADDk_vs_r_k_ARL__3D_mu02} assume $\mu=0.2$ while Figures~\ref{fig:ADDk_vs_r_k_ARL__3D_mu05} assume $\mu=0.5$. With regard to the level $\gamma>1$ of the ARL to false alarm, Figures~\ref{fig:ADDk_vs_r_k__3D_mu02_ARL100} and~\ref{fig:ADDk_vs_r_k__3D_mu05_ARL100} assume $\gamma=100$, Figures~\ref{fig:ADDk_vs_r_k__3D_mu02_ARL500} and~\ref{fig:ADDk_vs_r_k__3D_mu05_ARL500} are for $\gamma=500$, and Figures~\ref{fig:ADDk_vs_r_k__3D_mu02_ARL1000} and~\ref{fig:ADDk_vs_r_k__3D_mu05_ARL1000} assume $\gamma=1\,000$. There are two important observations to make from either set of figures. First, it is evident that giving the SR chart a positive headstart equips the chart with the Fast Initial Response (FIR) feature: the chart becomes more sensitive to initial out-of-control situations. However, the flip side of the FIR feature is that the chart gets slower in situations when the process is initially in control but goes out of control later. It is worth reiterating that in order to retain the level of the ARL to false alarm assigning a higher value to the headstart is offset by an appropriate upward adjustment of the control limit. The second observation is that the maximal ADD, i.e., $\SADD(\mathcal{S}_{A}^{r})\triangleq\max_{0\le k<\infty}\ADD_k(\mathcal{S}_{A}^{r})$, is a sophisticated function of $r$, and the specific value of $k$ at which the maximum is attained is hard to predict. As an aside, it is worth pointing out that the convergence of the ADDs to the steady-state regime is faster for $\mu=0.5$ than for $\mu=0.2$, which is consistent with one's intuition.

\begin{sidewaysfigure}[p]
    \centering
    \subfloat[$\gamma=100$.]{
        \includegraphics[width=0.33\textwidth]{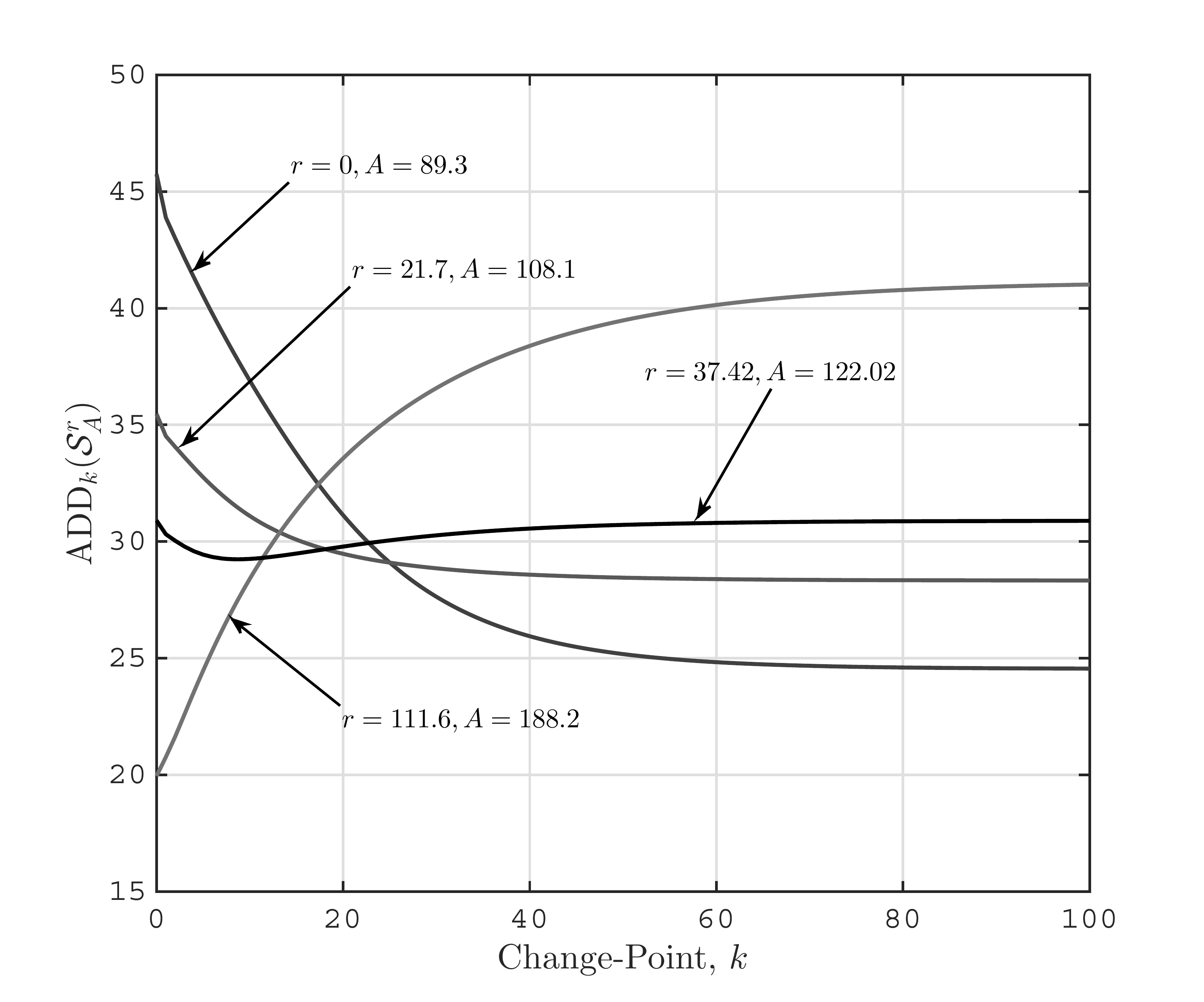}
    } %
    \subfloat[$\gamma=500$.]{
        \includegraphics[width=0.33\textwidth]{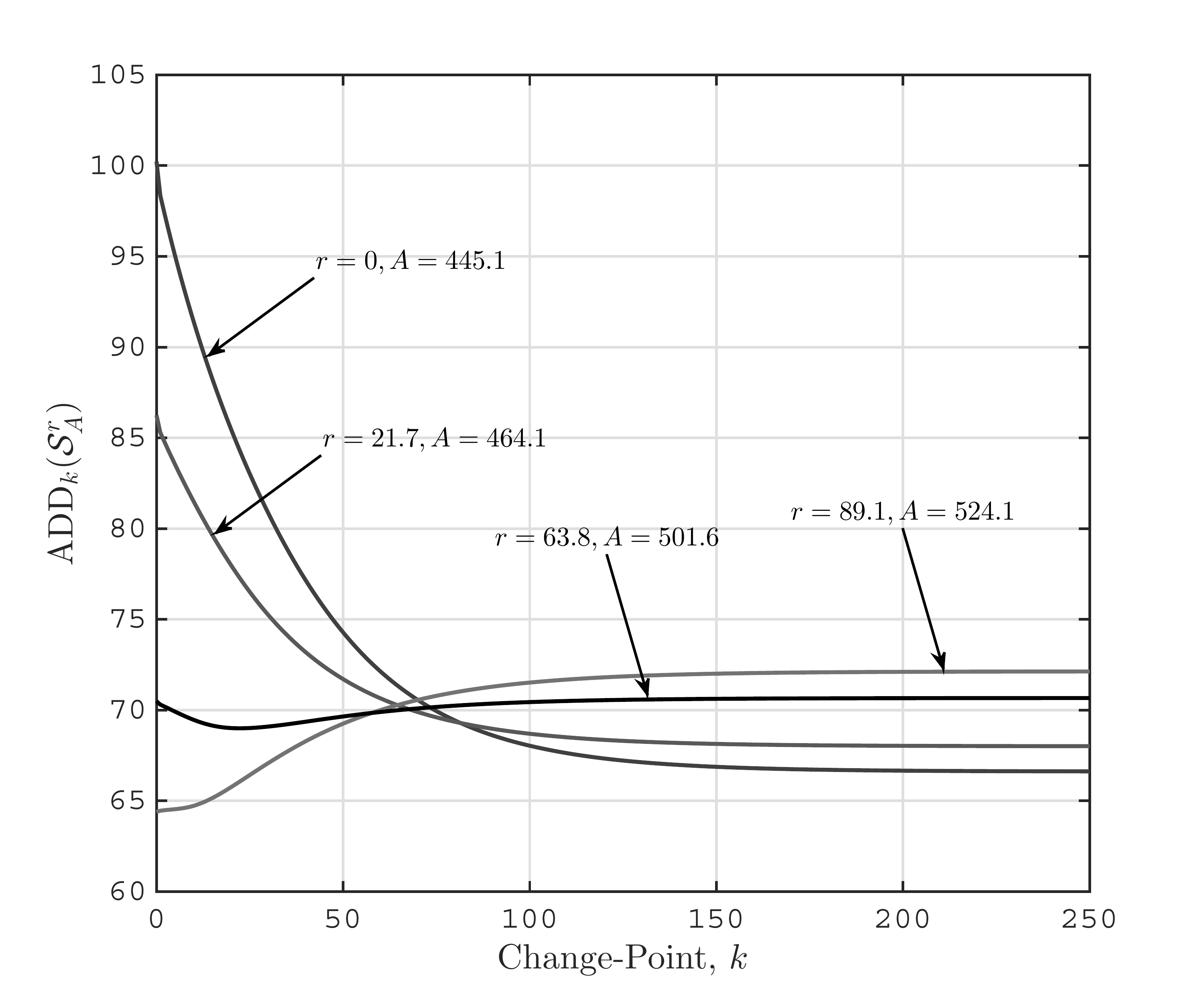}
    } %
    \subfloat[$\gamma=1\,000$.]{
        \includegraphics[width=0.33\textwidth]{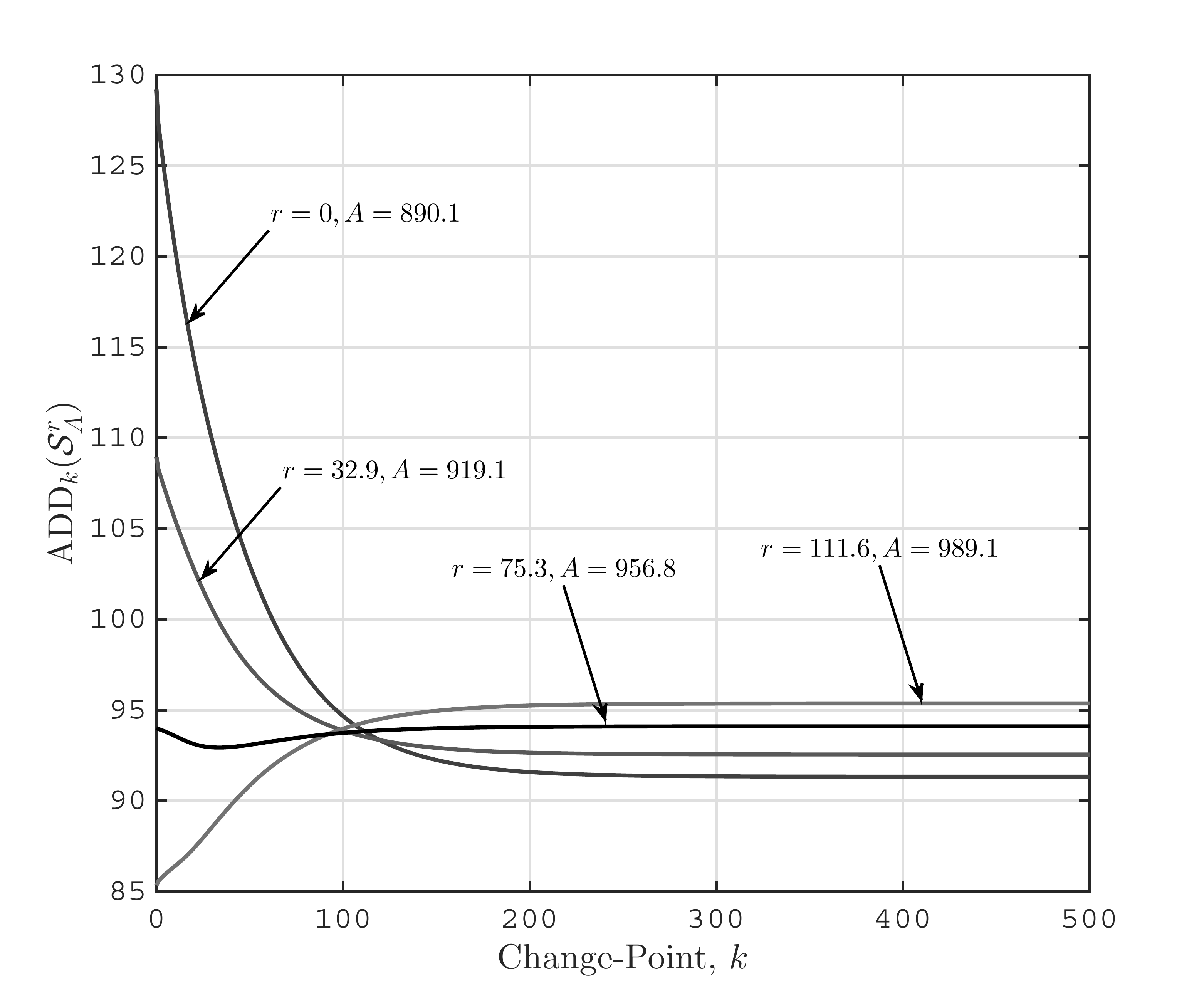}
    } %
    \caption{$\ADD_k(\mathcal{S}_{A}^{r})$ as a function of the headstart $R_0^r=r\ge0$, the change-point $k=0,1,\ldots$, and the ARL to false alarm level $\ARL(\mathcal{S}_{A}^{r})=\gamma>1$ for $\mu=0.2$.}
    \label{fig:ADDk_vs_r_k_ARL__mu02}
\end{sidewaysfigure}

\begin{sidewaysfigure}[p]
    \centering
    \subfloat[$\gamma=100$.]{
        \includegraphics[width=0.33\textwidth]{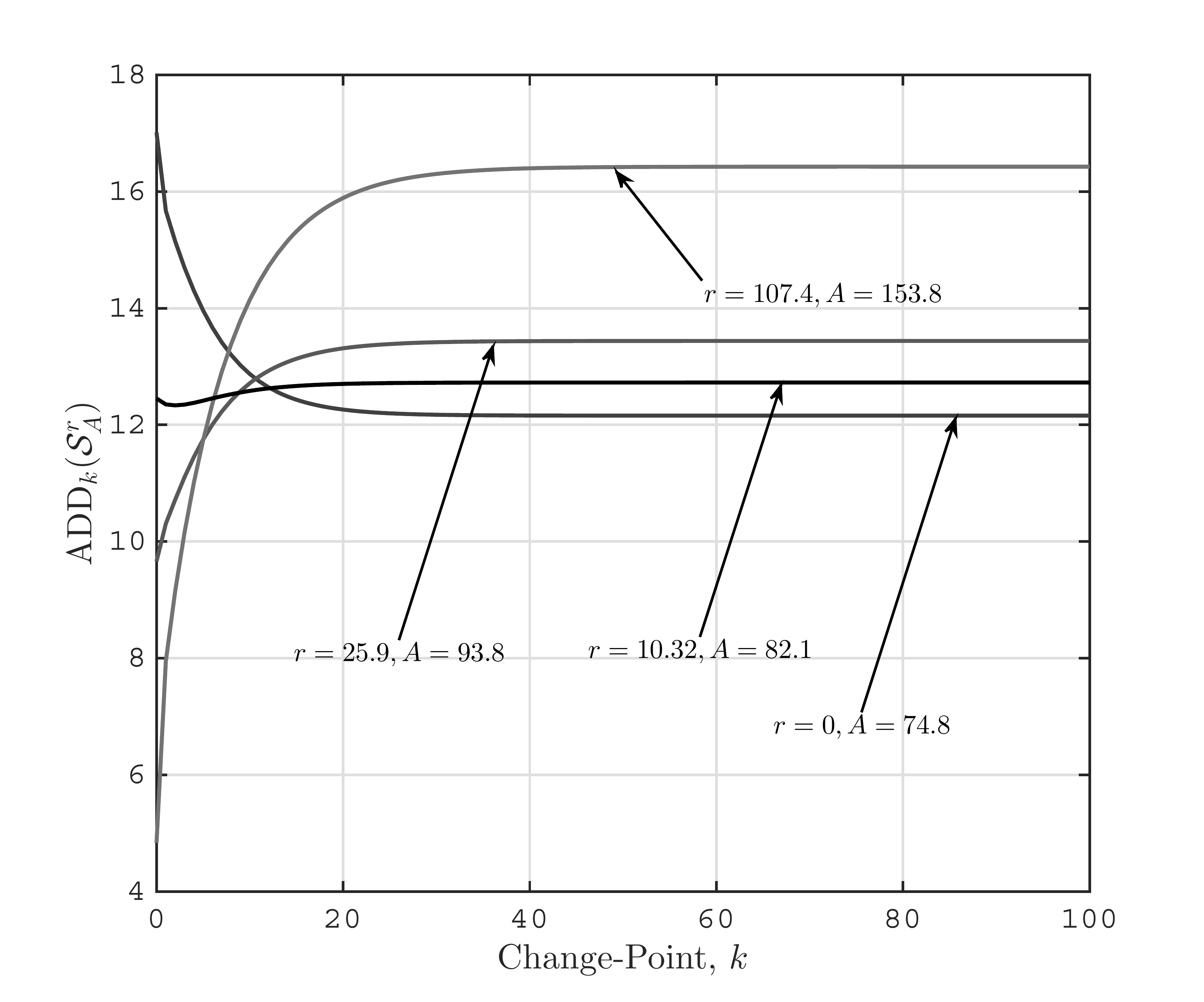}
    } %
    \subfloat[$\gamma=500$.]{
        \includegraphics[width=0.33\textwidth]{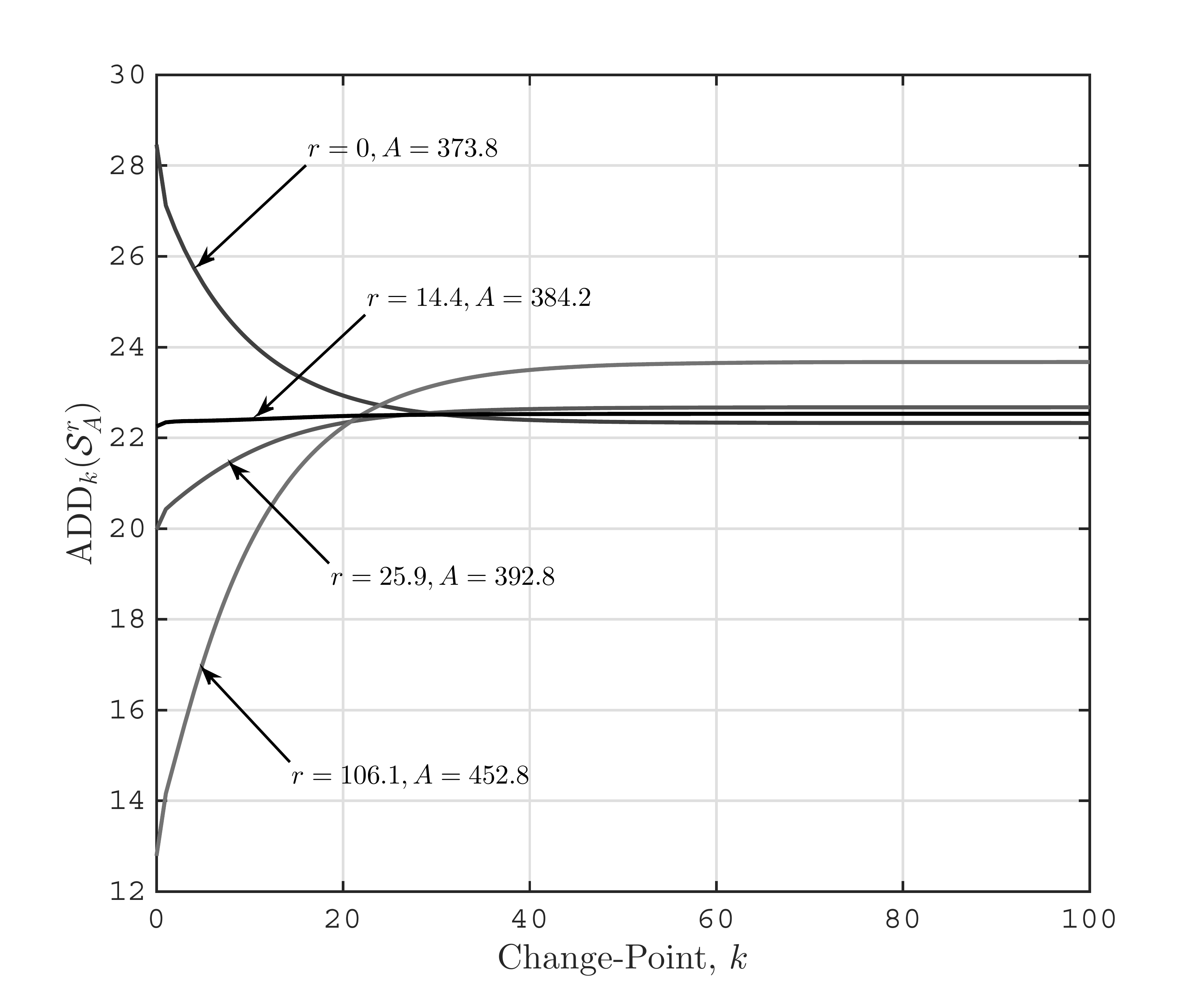}
    } %
    \subfloat[$\gamma=1\,000$.]{
        \includegraphics[width=0.33\textwidth]{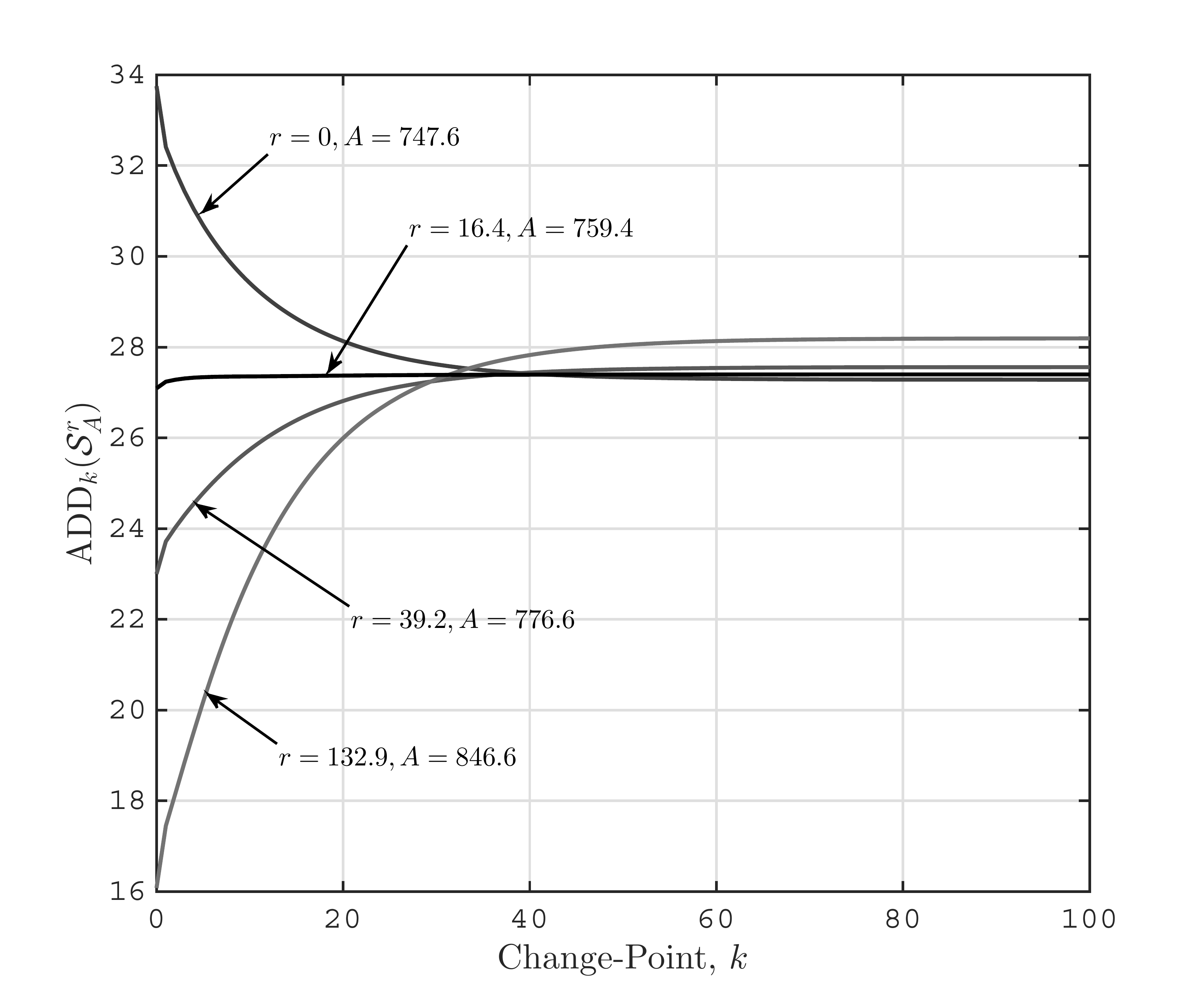}
    } %
    \caption{$\ADD_k(\mathcal{S}_{A}^{r})$ as a function of the headstart $R_0^r=r\ge0$, the change-point $k=0,1,\ldots$, and the ARL to false alarm level $\ARL(\mathcal{S}_{A}^{r})=\gamma>1$ for $\mu=0.5$.}
    \label{fig:ADDk_vs_r_k_ARL__mu05}
\end{sidewaysfigure}

\begin{sidewaysfigure}[p]
    \centering
    \subfloat[$\gamma=100$.]{
        \includegraphics[width=0.33\textwidth]{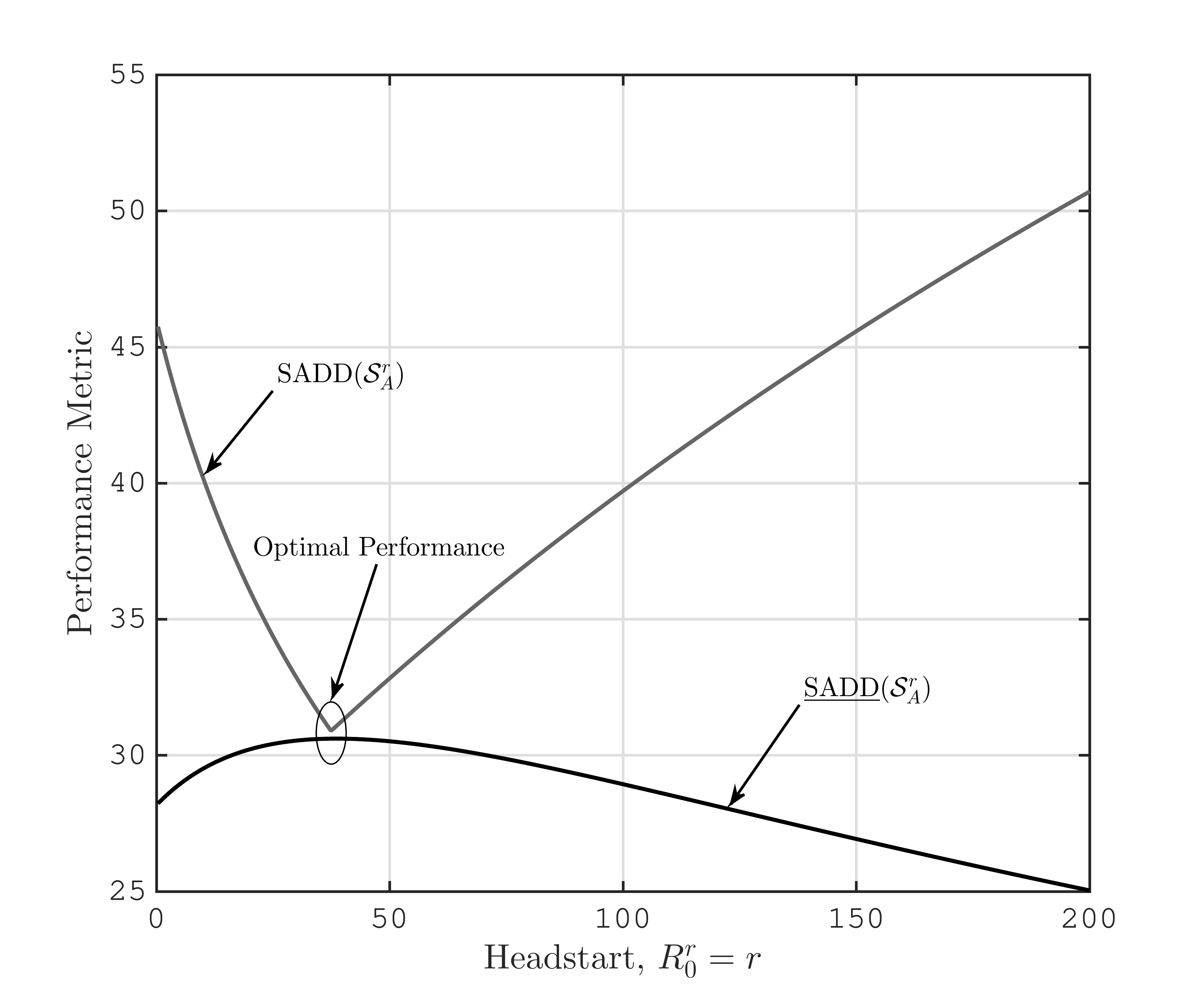}
    } %
    \subfloat[$\gamma=500$.]{
        \includegraphics[width=0.33\textwidth]{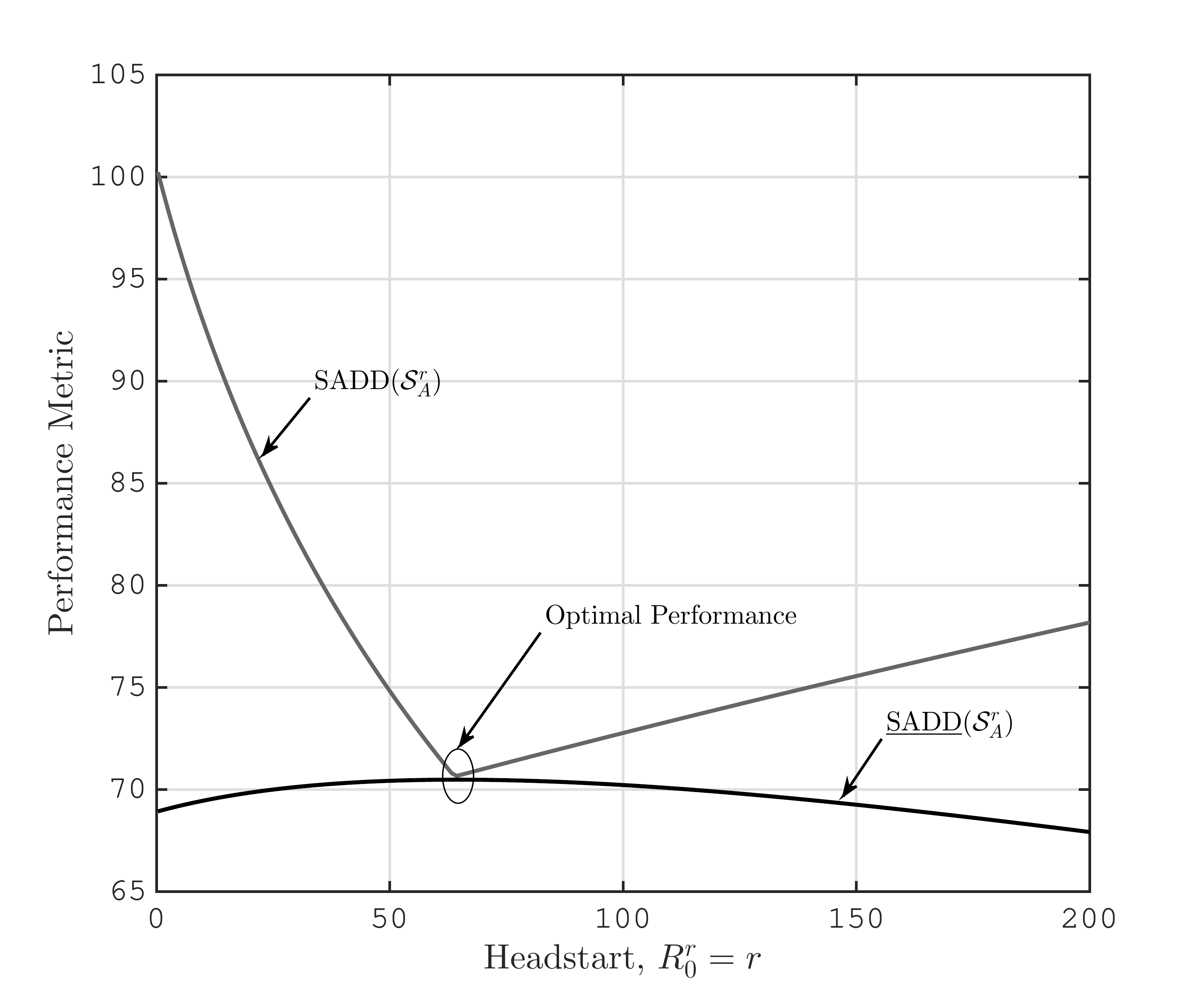}
    } %
    \subfloat[$\gamma=1\,000$.]{
        \includegraphics[width=0.33\textwidth]{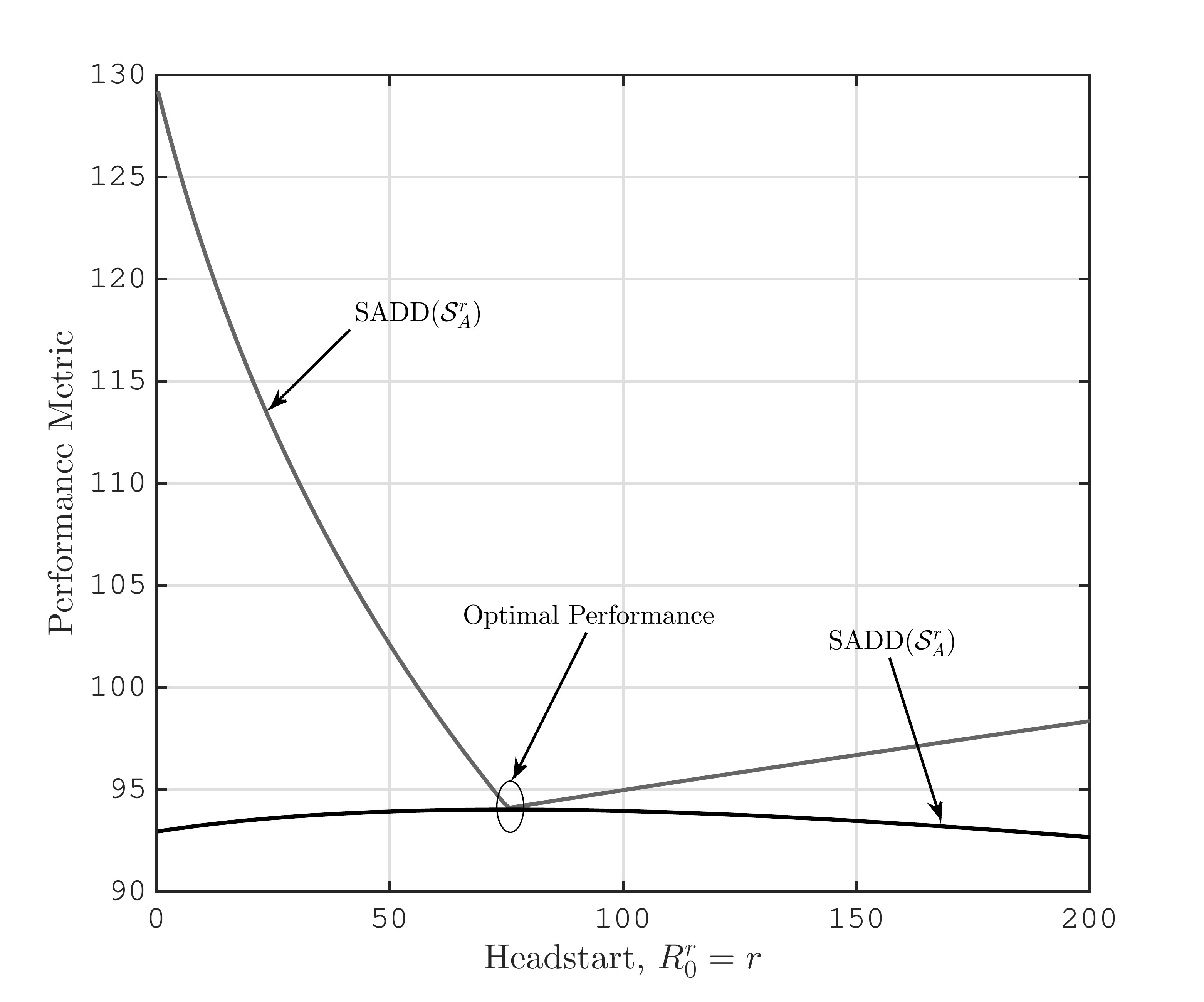}
    } %
    \caption{$\SADD(\mathcal{S}_{A}^{r})$ and $\underline{\SADD}(\mathcal{S}_{A}^{r})$ as functions of the headstart $R_0^r=r$ and the ARL to false alarm level $\ARL(\mathcal{S}_{A}^{r})=\gamma>1$ for $\mu=0.2$.}
    \label{fig:SADD_LwrBnd_vs_r_ARL__mu02}
\end{sidewaysfigure}

\begin{sidewaysfigure}[p]
    \centering
    \subfloat[$\gamma=100$.]{\label{fig:SADD_LwrBnd_vs_r__mu05_ARL100}
        \includegraphics[width=0.33\textwidth]{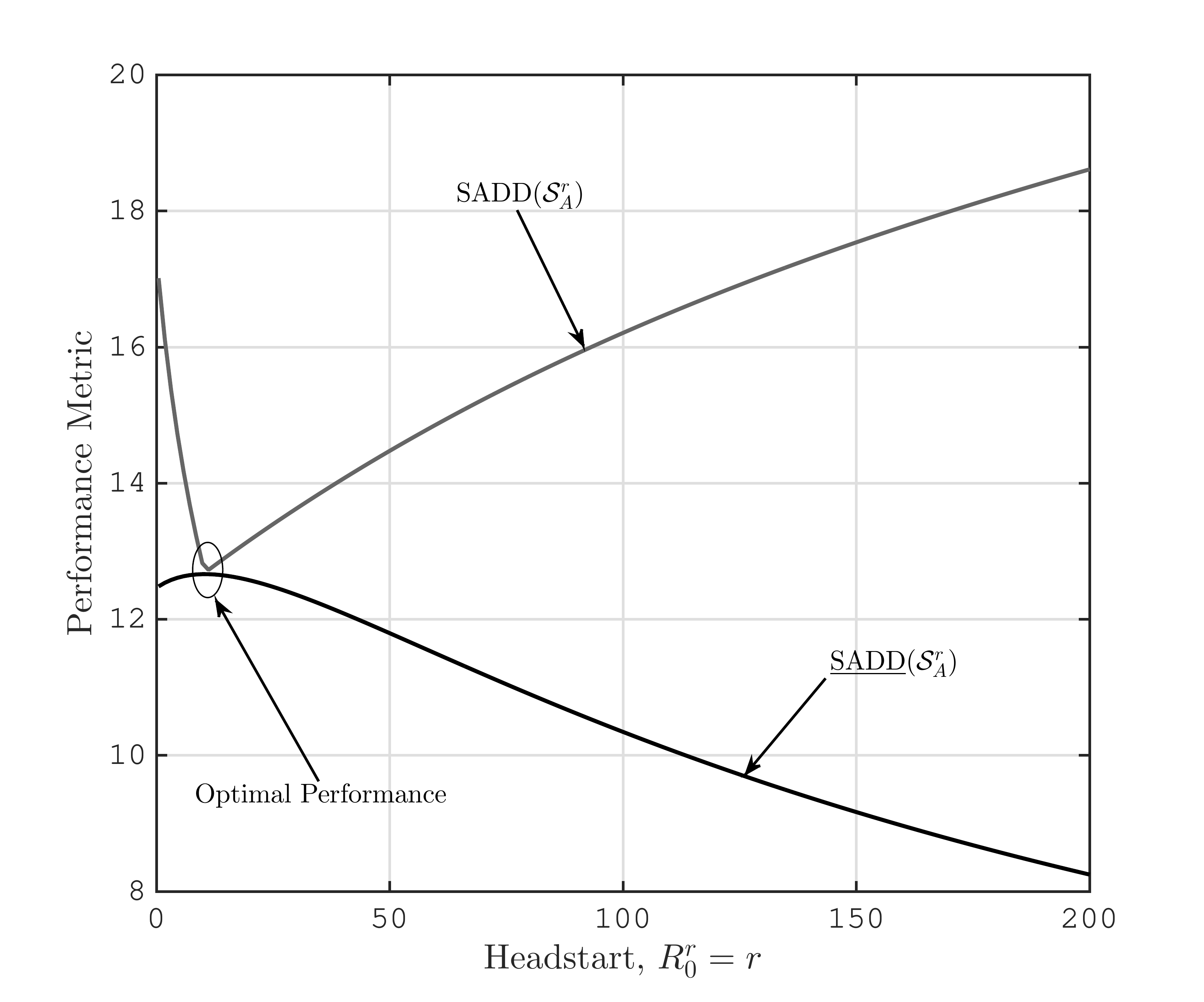}
    } %
    \subfloat[$\gamma=500$.]{\label{fig:SADD_LwrBnd_vs_r__mu05_ARL500}
        \includegraphics[width=0.33\textwidth]{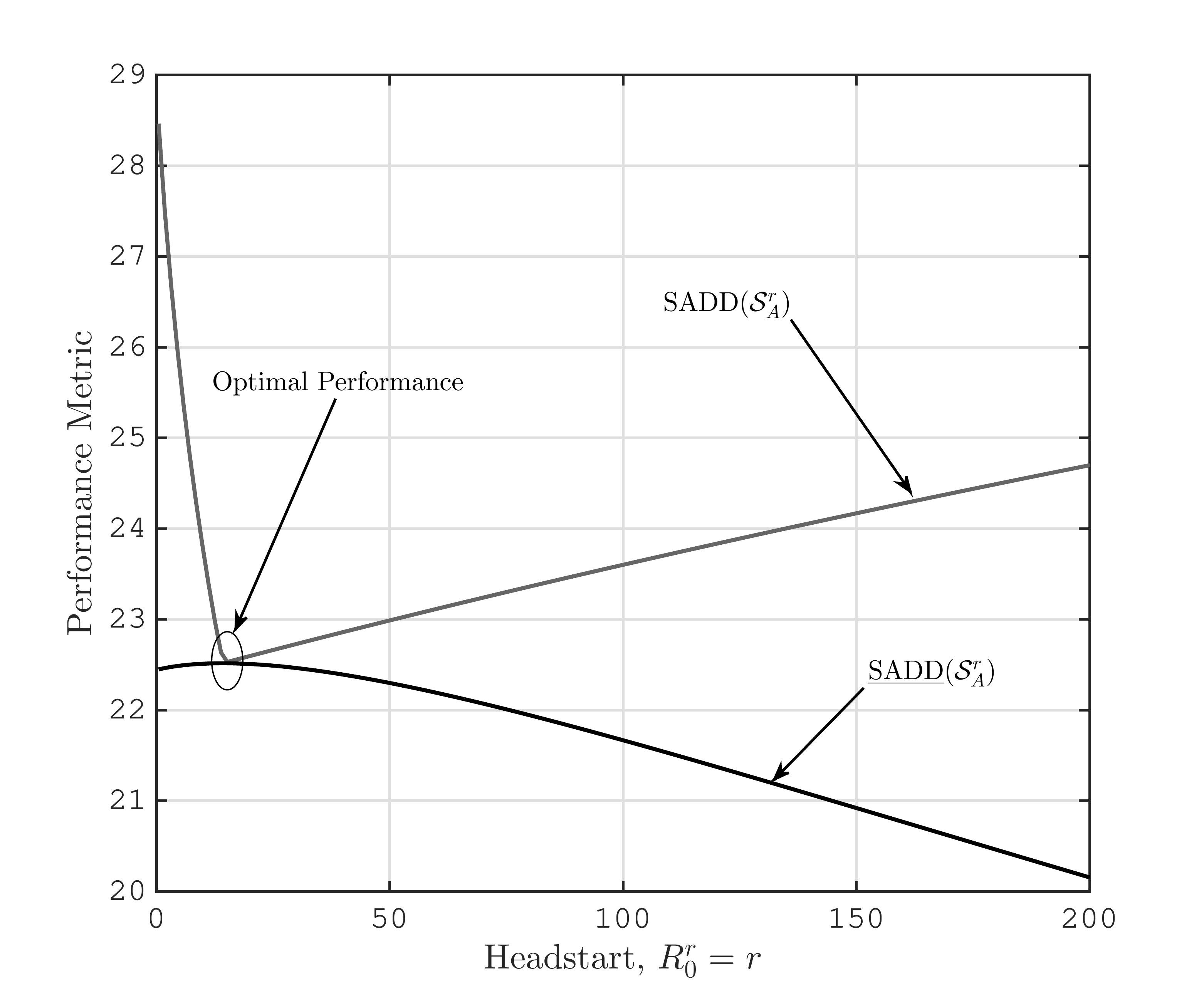}
    } %
    \subfloat[$\gamma=1\,000$.]{\label{fig:SADD_LwrBnd_vs_r__mu05_ARL1000}
        \includegraphics[width=0.33\textwidth]{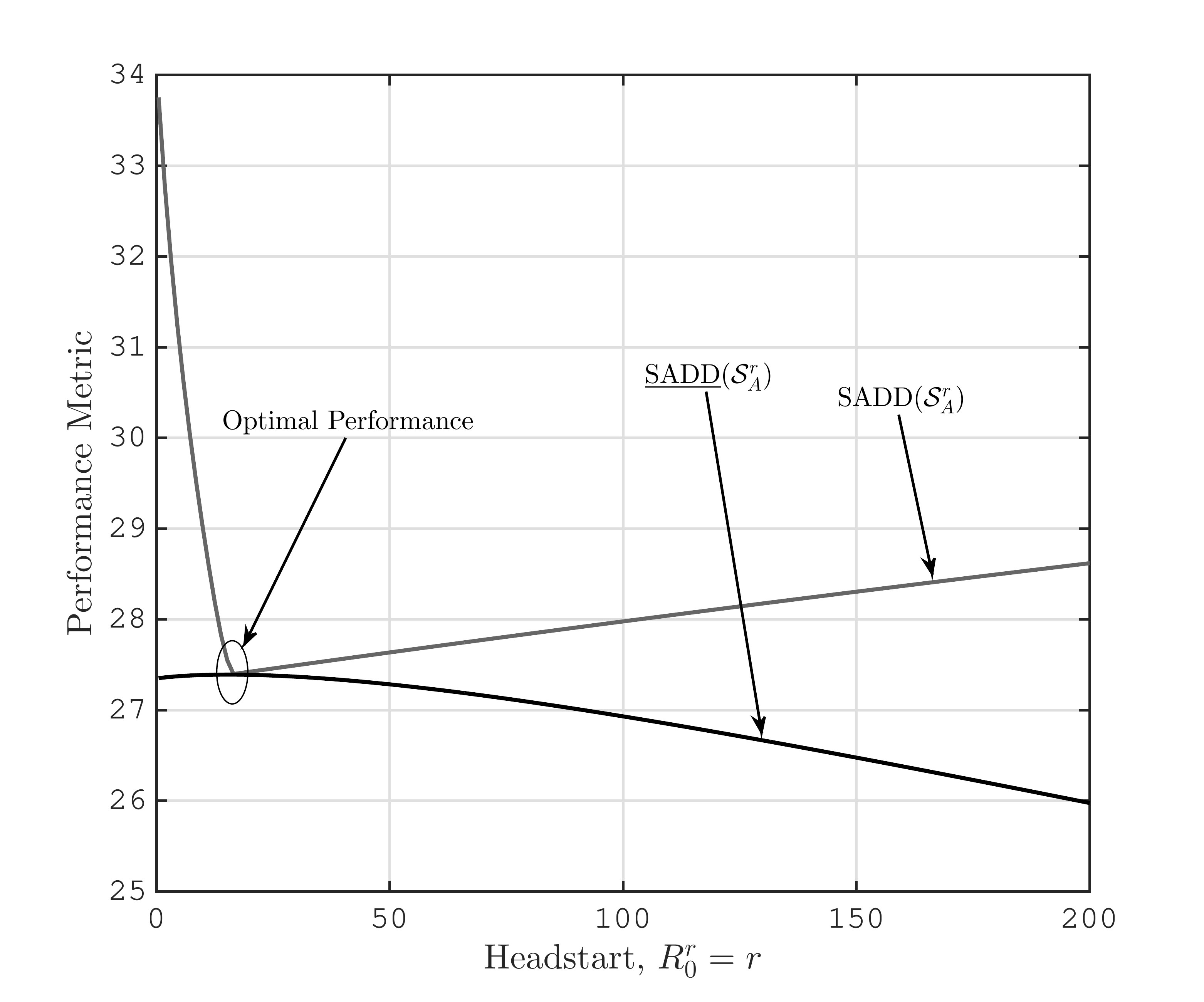}
    } %
    \caption{$\SADD(\mathcal{S}_{A}^{r})$ and $\underline{\SADD}(\mathcal{S}_{A}^{r})$ as functions of the headstart $R_0^r=r$ and the ARL to false alarm level $\ARL(\mathcal{S}_{A}^{r})=\gamma>1$ for $\mu=0.5$.}
    \label{fig:SADD_LwrBnd_vs_r_ARL__mu05}
\end{sidewaysfigure}

To better illustrate the FIR feature at work, let us look at Figures~\ref{fig:ADDk_vs_r_k_ARL__mu02} and~\ref{fig:ADDk_vs_r_k_ARL__mu05}, which are effectively the projections of the 3D surfaces shown in Figures~\ref{fig:ADDk_vs_r_k_ARL__3D_mu02} and~\ref{fig:ADDk_vs_r_k_ARL__3D_mu05} onto the $(k, \ADD_k(\mathcal{S}_{A}^{r}))$-plane, made for a selection of values of $r$. Specifically, Figures~\ref{fig:ADDk_vs_r_k_ARL__mu02} assume $\mu=0.2$ and Figures~\ref{fig:ADDk_vs_r_k_ARL__mu02} are for $\mu=0.5$. The corresponding levels $\gamma>1$ of the ARL to false alarm are given in the figures' subtitles. The figures clearly demonstrate that, as the headstart increases, the performance of the GSR chart for initial of early out-of-control situation improves. However, the performance in situations when the process goes out of control later degrades. The interesting question is whether it is possible to optimize this tradeoff. This question is hard to answer properly without getting the lowerbound $\SADD(\mathcal{S}_{A}^{r})$ involved, as is done in Figures~\ref{fig:SADD_LwrBnd_vs_r_ARL__mu02} and~\ref{fig:SADD_LwrBnd_vs_r_ARL__mu05}.

Specifically, Figures~\ref{fig:SADD_LwrBnd_vs_r_ARL__mu02} and~\ref{fig:SADD_LwrBnd_vs_r_ARL__mu05} provide an idea as to the manner in which $\SADD(\mathcal{S}_A^r)$ and $\underline{\SADD}(\mathcal{S}_A^r)$ each depend on the headstart, assuming, as before, that every change in the headstart is accompanied by the appropriate adjustment of the detection threshold, so that the ARL to false alarm constraint is kept intact. More specifically, Figures~\ref{fig:SADD_LwrBnd_vs_r_ARL__mu02} correspond to $\mu=0.2$ and Figures~\ref{fig:SADD_LwrBnd_vs_r_ARL__mu05} are for $\mu=0.5$. The respective levels $\gamma$ of the ARL to false alarm are again given in the subtitles.

It is evident from the figures that, regardless of the contrastness of the shift in the mean $\mu\neq0$ and no matter the ARL to false alarm level $\gamma>1$, the lowerbound is an upward arching smooth function of the initial score, and it has a distinct maximum. The figures also clearly indicate that the maximal ADD as a function of $r$ has a minimum with the appearance of a down pointing cusp; the cusp is an indication that the way the maximal element of the sequence $\{\ADD_{k}(\mathcal{S}_A^r)\}_{k\ge 0}$ and its location within the sequence depend on the headstart is highly nonlinear. The essential observation is that the lowerbound appears to peak at approximately the same (slightly smaller actually) headstart value as that at which the maximal ADD is minimized. Moreover, although the maximal ADD's minimum is higher than the lowerbound's maximum, the difference is not practically significant, even if $\gamma$ is as low as $100$, and is smaller, the higher the value of $\gamma$. Therefore, any other chart with the same level of the ARL to false alarm cannot possibly detect the shift in the mean with a detection delay substantially lower than that delivered by the optimized GSR chart, especially if the shift in the mean is contrast.

To draw a line under this section, in Tables~\ref{tab:opt_SADD_r_A_LwrBnd_vs_theta_gamma_p1} and~\ref{tab:opt_SADD_r_A_LwrBnd_vs_theta_gamma_p2}, we give the optimal headstart and detection threshold values that have been computed by solving the constrained optimization problem~\eqref{eq:opt-r-A-choice} for $\gamma=\{100,200,\ldots,900,1\,000\}$ and $\mu=\{0.1,0.2,\ldots,0.9,1.0\}$. Recall also that our data model is symmetric with respect to the sign of $\mu\neq0$. The tables also include the corresponding $\SADD(\mathcal{S}_{A}^{r})$ and $\underline{\SADD}(\mathcal{S}_{A}^{r})$ values. One can see from the tables that $\SADD(\mathcal{S}_{A}^{r})\approx\underline{\SADD}(\mathcal{S}_{A}^{r})$, which is to say that the detection capabilities of the optimized GSR chart are almost the best. One can also see that the effect of headstarting is the stronger, the fainter the anticipated shift in the mean. If the latter is fairly contrast, the optimal headstart value is close to zero. In addition, the tables suggest that the optimal headstart value, as a function of the ARL to false alarm level $\gamma>1$, has a {\em finite} limit as $\gamma\to+\infty$; the convergence to the limiting value is the slower, the weaker the change. However, a closed-form formula for this limiting value is prohibitively difficult to obtain.
\begin{sidewaystable}[p]
    \centering
    \caption{Optimal headstart, $r^{*}>0$, control limit, $A^{*}>0$, maximal ADD, $\SADD(\mathcal{S}_{A}^{r})$, and the lowerbound, $\underline{\SADD}(\mathcal{S}_{A}^{r})$, as functions of the shift magnitude, $\mu>0$, and the ARL to false alarm level, $\ARL(\mathcal{S}_{A}^{r})=\gamma>1$, for $\gamma=\{100, 200, 300, 400, 500\}$.}
    \setlength\extrarowheight{1pt}
    \begin{tabular}{@{}|c||c||*{10}{c|}@{}}
    \hline
      \multirow{2}{*}{$\ARL(\mathcal{S}_{A}^{r})=\gamma$} & {\bfseries Performance} & \multicolumn{10}{c|}{\bfseries Change Magnitude ($\mu>0$)}\\
    \cline{3-12}
      & {\bfseries Characteristic} & 0.1 & 0.2 & 0.3 & 0.4 & 0.5 & 0.6 & 0.7 & 0.8 & 0.9 & 1.0\\
        \hline\hline\multirow{4}{*}{$100$}
                     & $r^{*}$& $83.93$& $37.42$& $21.96$& $14.53$& $10.32$& $7.66$& $5.89$& $4.64$& $3.74$& $3.05$ \\
        \cline{2-12} & $A^{*}$& $173.25$& $122.02$& $102.11$& $90.43$& $82.14$& $75.6$& $70.14$& $65.38$& $61.14$& $57.31$ \\
        \cline{2-12} & $\SADD(\mathcal{S}_{A}^{r})$& $49.65$& $30.9$& $21.6$& $16.17$& $12.68$& $10.28$& $8.55$& $7.25$& $6.25$& $5.46$  \\
        \cline{2-12} & $\underline{\SADD}(\mathcal{S}_{A}^{r})$& $48.76$& $30.62$& $21.49$& $16.13$& $12.66$& $10.27$& $8.54$& $7.25$& $6.25$& $5.46$ \\
    \hline\hline\multirow{4}{*}{$200$}
                     & $r^{*}$& $114.43$& $48.29$& $27.23$& $17.49$& $12.14$& $8.86$& $6.72$& $5.27$& $4.24$& $3.48$ \\
        \cline{2-12} & $A^{*}$& $296.37$& $220.71$& $190.51$& $172$& $158.26$& $147$& $137.28$& $128.63$& $120.79$& $113.58$ \\
        \cline{2-12} & $\SADD(\mathcal{S}_{A}^{r})$& $79.79$& $45.39$& $30.1$& $21.75$& $16.61$& $13.19$& $10.79$& $9.03$& $7.69$& $6.65$  \\
        \cline{2-12} & $\underline{\SADD}(\mathcal{S}_{A}^{r})$& $78.7$& $45.14$& $30.03$& $21.72$& $16.6$& $13.19$& $10.79$& $9.03$& $7.69$& $6.65$ \\
    \hline\hline\multirow{4}{*}{$300$}
                     & $r^{*}$& $135.53$& $55.1$& $30.29$& $19.12$& $13.1$& $9.54$& $7.27$& $5.7$& $4.6$& $3.77$ \\
        \cline{2-12} & $A^{*}$& $410.61$& $315.77$& $277.06$& $252.52$& $233.74$& $218.04$& $204.24$& $191.76$& $180.34$& $169.78$ \\
        \cline{2-12} & $\SADD(\mathcal{S}_{A}^{r})$& $103.23$& $55.71$& $35.87$& $25.41$& $19.13$& $15.04$& $12.19$& $10.13$& $8.58$& $7.38$  \\
        \cline{2-12} & $\underline{\SADD}(\mathcal{S}_{A}^{r})$& $102.08$& $55.5$& $35.81$& $25.4$& $19.13$& $15.03$& $12.19$& $10.13$& $8.58$& $7.38$ \\
    \hline\hline\multirow{4}{*}{$400$}
                     & $r^{*}$& $151.87$& $60.02$& $32.41$& $20.2$& $13.81$& $10.05$& $7.65$& $6.01$& $4.83$& $3.98$ \\
        \cline{2-12} & $A^{*}$& $520.37$& $409.15$& $362.81$& $332.61$& $309.04$& $288.95$& $271.07$& $254.81$& $239.82$& $225.94$ \\
        \cline{2-12} & $\SADD(\mathcal{S}_{A}^{r})$& $122.8$& $63.86$& $40.29$& $28.17$& $21.02$& $16.4$& $13.22$& $10.93$& $9.22$& $7.91$  \\
        \cline{2-12} & $\underline{\SADD}(\mathcal{S}_{A}^{r})$& $121.65$& $63.68$& $40.25$& $28.16$& $21.01$& $16.39$& $13.22$& $10.93$& $9.22$& $7.91$ \\
    \hline\hline\multirow{4}{*}{$500$}
                     & $r^{*}$& $165.27$& $63.84$& $33.98$& $21.03$& $14.36$& $10.45$& $7.95$& $6.25$& $5.03$& $4.14$ \\
        \cline{2-12} & $A^{*}$& $627.35$& $501.56$& $448.1$& $412.5$& $384.21$& $359.78$& $337.86$& $317.81$& $299.29$& $282.07$ \\
        \cline{2-12} & $\SADD(\mathcal{S}_{A}^{r})$& $139.75$& $70.63$& $43.86$& $30.39$& $22.52$& $17.48$& $14.05$& $11.57$& $9.73$& $8.33$  \\
        \cline{2-12} & $\underline{\SADD}(\mathcal{S}_{A}^{r})$& $138.63$& $70.48$& $43.86$& $30.39$& $22.52$& $17.48$& $14.03$& $11.57$& $9.73$& $8.33$ \\
    \hline
    \end{tabular}
    \label{tab:opt_SADD_r_A_LwrBnd_vs_theta_gamma_p1}
\end{sidewaystable}
\begin{sidewaystable}[p]
    \centering
    \caption{Optimal headstart, $r^{*}>0$, control limit, $A^{*}>0$, maximal ADD, $\SADD(\mathcal{S}_{A}^{r})$, and the lowerbound, $\underline{\SADD}(\mathcal{S}_{A}^{r})$, as functions of the shift magnitude, $\mu>0$, and the ARL to false alarm level, $\ARL(\mathcal{S}_{A}^{r})=\gamma>1$, for $\gamma=\{600, 700, 800, 900, 1\,000\}$.}
    \setlength\extrarowheight{1pt}
    \begin{tabular}{@{}|c||c||*{10}{c|}@{}}
    \hline
      \multirow{2}{*}{$\ARL(\mathcal{S}_{A}^{r})=\gamma$} & {\bfseries Performance} & \multicolumn{10}{c|}{\bfseries Change Magnitude ($\mu>0$)}\\
    \cline{3-12}
      & {\bfseries Characteristic} & 0.1 & 0.2 & 0.3 & 0.4 & 0.5 & 0.6 & 0.7 & 0.8 & 0.9 & 1.0\\
    \hline\hline\multirow{4}{*}{$600$}
                     & $r^{*}$& $176.63$& $66.94$& $35.24$& $21.73$& $14.81$& $10.78$& $8.19$& $6.45$& $5.2$& $4.28$ \\
        \cline{2-12} & $A^{*}$& $732.41$& $593.32$& $533.12$& $492.28$& $459.31$& $430.56$& $404.61$& $380.8$& $358.73$& $338.18$ \\
        \cline{2-12} & $\SADD(\mathcal{S}_{A}^{r})$& $153.8$& $76.46$& $46.95$& $32.26$& $23.77$& $18.38$& $14.71$& $12.09$& $10.15$& $8.67$  \\
        \cline{2-12} & $\underline{\SADD}(\mathcal{S}_{A}^{r})$& $153.71$& $76.33$& $46.92$& $32.25$& $23.77$& $18.37$& $14.71$& $12.09$& $10.15$& $8.67$ \\
    \hline\hline\multirow{4}{*}{$700$}
                     & $r^{*}$& $186.49$& $69.53$& $36.25$& $22.32$& $15.21$& $11.06$& $8.42$& $6.61$& $5.34$& $4.39$ \\
        \cline{2-12} & $A^{*}$& $836.06$& $684.63$& $617.94$& $571.98$& $534.37$& $501.32$& $471.35$& $443.76$& $418.15$& $394.28$ \\
        \cline{2-12} & $\SADD(\mathcal{S}_{A}^{r})$& $168.37$& $81.58$& $49.59$& $33.87$& $24.85$& $19.15$& $15.29$& $12.54$& $10.51$& $8.96$  \\
        \cline{2-12} & $\underline{\SADD}(\mathcal{S}_{A}^{r})$& $167.32$& $81.47$& $49.57$& $33.86$& $24.85$& $19.14$& $15.28$& $12.54$& $10.51$& $8.96$ \\
    \hline\hline\multirow{4}{*}{$800$}
                     & $r^{*}$& $195.2$& $71.73$& $37.14$& $22.84$& $15.55$& $11.3$& $8.6$& $6.77$& $5.47$& $4.5$ \\
        \cline{2-12} & $A^{*}$& $938.62$& $775.59$& $702.67$& $651.62$& $609.38$& $572.04$& $538.06$& $506.71$& $477.58$& $450.38$ \\
        \cline{2-12} & $\SADD(\mathcal{S}_{A}^{r})$& $180.76$& $86.16$& $51.94$& $35.29$& $25.79$& $19.82$& $15.79$& $12.93$& $10.82$& $9.22$  \\
        \cline{2-12} & $\underline{\SADD}(\mathcal{S}_{A}^{r})$& $179.75$& $86.06$& $51.92$& $35.28$& $25.79$& $19.82$& $15.79$& $12.93$& $10.82$& $9.22$ \\
    \hline\hline\multirow{4}{*}{$900$}
                     & $r^{*}$& $202.99$& $73.65$& $37.93$& $23.31$& $15.86$& $11.53$& $8.78$& $6.91$& $5.57$& $4.59$ \\
        \cline{2-12} & $A^{*}$& $1\,040.31$& $866.3$& $787.3$& $731.22$& $684.37$& $642.75$& $604.77$& $569.66$& $536.98$& $506.46$ \\
        \cline{2-12} & $\SADD(\mathcal{S}_{A}^{r})$& $192.18$& $90.3$& $54.04$& $36.55$& $26.64$& $20.42$& $16.24$& $13.28$& $11.1$& $9.44$  \\
        \cline{2-12} & $\underline{\SADD}(\mathcal{S}_{A}^{r})$& $191.21$& $90.21$& $54.03$& $36.55$& $26.63$& $20.42$& $16.24$& $13.28$& $11.1$& $9.44$ \\
    \hline\hline\multirow{4}{*}{$1\,000$}
                     & $r^{*}$& $210.04$& $75.34$& $38.62$& $23.71$& $16.14$& $11.73$& $8.93$& $7.04$& $5.68$& $4.66$ \\
        \cline{2-12} & $A^{*}$& $1\,141.3$& $956.81$& $871.86$& $810.77$& $759.35$& $713.44$& $671.46$& $632.6$& $596.38$& $562.54$ \\
        \cline{2-12} & $\SADD(\mathcal{S}_A^r)$& $202.79$& $94.09$& $55.96$& $37.7$& $27.39$& $20.96$& $16.64$& $13.59$& $11.35$& $9.65$  \\
        \cline{2-12} & $\underline{\SADD}(\mathcal{S}_{A}^{r})$& $201.86$& $94.01$& $55.94$& $37.69$& $27.39$& $20.95$& $16.64$& $13.59$& $11.35$& $9.64$ \\
    \hline
    \end{tabular}
    \label{tab:opt_SADD_r_A_LwrBnd_vs_theta_gamma_p2}
\end{sidewaystable}

\section{Concluding Remarks}\label{c12:sec:conclusion}
In summary we see that
\begin{enumerate}
    \setlength{\itemsep}{3pt}
    \item Starting an SR chart off a nonzero initial score lessens the ARL to false alarm, so that the chart's in-control performance is worse than when no headstart is used. On the flip side, however, the chart becomes more sensitive to initial out-of-control situations. This is precisely the FIR phenomenon.
    \item The drop in the ARL to false alarm caused by a positive headstart value can be compensated by an increase of the control limit. While this would negatively affect the chart's out-of-control performance, the magnitude of the effect appears to be not substantial.
    \item The FIR feature comes at the price of poorer performance in situations when the process under surveillance is initially in control but goes out of control later. In particular, if the process is not expected to shift out of control for a long while, then no headstarting is necessary, because the SR chart's steady-state performance would degrade otherwise.
\end{enumerate}

The same observations were previously made by Lucas \& Crosier~\citeyearpar{Lucas+Crosier:T1982} about the CUSUM chart.

Our additional and more important contribution consists in a deeper investigation of the headstart-vs-control-limit tradeoff: the overall performance of the GSR chart optimized not only with respect to the headstart but also with respect to the control limit proved to be nearly the best one can get amid complete uncertainty as to when the observed process may go out of control. This is a direct implication of the GSR chart's strong optimality properties established by~\cite{Pollak+Tartakovsky:SS2009}, \cite{Shiryaev+Zryumov:Khabanov2010}, \cite{Tartakovsky+Polunchenko:IWAP2010}, \cite{Polunchenko+Tartakovsky:AS2010}, and by~\cite{Tartakovsky+etal:TPA2012}. The optimal headstart and control limit values, and the corresponding out-of-control performance and its lowerbound, for a variety of cases, are given in Tables~\ref{tab:opt_SADD_r_A_LwrBnd_vs_theta_gamma_p1} and~\ref{tab:opt_SADD_r_A_LwrBnd_vs_theta_gamma_p2}.

The benefits of optimizing the GSR chart are the greater, the fainter the change. From a practical standpoint, this means that if one is interested in detecting a faint change, then the GSR chart with optimally selected control limit and headstart is the way to go. The size of the actual efficiency improvement can be estimated using Tables~\ref{tab:opt_SADD_r_A_LwrBnd_vs_theta_gamma_p1} and~\ref{tab:opt_SADD_r_A_LwrBnd_vs_theta_gamma_p2}. However, if the anticipated change to be detected is more or less contrast, then the GSR chart, whether optimized or not, will not offer any substantial advantage (in terms of the speed of detection) over the CUSUM scheme or the EWMA chart.

\begin{acknowledgement}
The author's effort was partially supported by the Simons Foundation via a Collaboration Grant in Mathematics under Award \#\,304574.
\end{acknowledgement}


\end{document}